\documentclass[aps, pre, reprint, superscriptaddress,hyperref, onecolumn, showkeys]{revtex4-2}
\usepackage[pdftex]{graphicx}
\graphicspath{{./Figures}} %Setting the graphicspath
\usepackage{dcolumn} 
\usepackage{bm} 
\usepackage[pdftex]{hyperref} 
\usepackage{longtable} 
\usepackage[T1]{fontenc}
\usepackage{times}
\usepackage{xcolor}
\usepackage{wrapfig}
\usepackage{float}
\usepackage[symbol]{footmisc}
\usepackage{amssymb,amsfonts,amsmath}
\usepackage{svg}
\usepackage{siunitx}
 \usepackage[pagewise]{lineno}
\usepackage{soul}
\DeclareSIUnit\angstrom{\text {Å}}

\renewcommand{\BibitemShut}[1]{}

\begin{document}

\title{Self-assembly of Colloidal Superballs Under Spherical Confinement of a Drying Droplet}

\author{Sarah\ \surname{Schyck}*}
\affiliation{Delft University of Technology,Department of Chemical Engineering, 2629 HZ Delft, The Netherlands}

\author{Janne-Mieke\ \surname{Meijer}*}
\affiliation{Eindhoven University of Technology, Department of Applied Physics and Insitute for Complex Molecular Systems, Eindhoven, The Netherlands}

\author{Lucia\ \surname{Baldauf}}
\affiliation{University of Amsterdam, Institute of Physics, Science Park 904, Amsterdam, The Netherlands}
\affiliation{Present address: Delft University of Technology,Department of Bionanoscience, Kavli Institute of Nanoscience, Delft 2629 HZ Delft, The Netherlands}

\author{Peter\ \surname{Schall}}
\affiliation{University of Amsterdam, Institute of Physics, Science Park 904, Amsterdam, The Netherlands}

\author{Andrei V.\ \surname{Petukhov}}
\affiliation{Utrecht University, Debye Institute for Nanomaterials Science, Utrecht, the Netherlands}
\affiliation{Eindhoven University of Technology, Department of Chemical Engineering and Chemistry, Eindhoven, the Netherlands}

\author{Laura\ \surname{Rossi}}
\email{L.Rossi@tudelft.nl}
\affiliation{Delft University of Technology,Department of Chemical Engineering, 2629 HZ Delft, The Netherlands}

\date{\today}

\begin{abstract}

Understanding the relationship between colloidal building block shape and self-assembled material structure is important for the development of novel materials by self-assembly. In this regard, colloidal superballs are unique building blocks because their shape can smoothly transition between spherical and cubic. Assembly of colloidal superballs under spherical confinement results in macroscopic clusters with ordered internal structure. By utilizing Small Angle X-Ray Scattering (SAXS), we probe the internal structure of colloidal superball dispersion droplets during confinement. We observe and identify four distinct drying regimes that arise during compression via evaporating droplets, and we track the development of the assembled macrostructure. As the superballs assemble, we found that they arrange into the predicted paracrystalline, rhombohedral C$_1$-lattice that varies by the constituent superballs' shape. This provides insights in the behavior between confinement and particle shape that can be applied in the development of new functional materials.  

\end{abstract}

\keywords{Small angle X-ray scattering; Colloidal superballs; Core-shell particles; Spherical confinement}

\maketitle

\linespread{1.6}\selectfont{}

The self-assembly of colloidal particles is an ideal mechanism for structuring matter at the nano- and microscale, and can produce materials of interest for different applications, such as photonics, opto-electronics, catalysis or energy storage \cite{Manoharan2015, Zhao2020}. 
To direct the self-assembly process towards the desired structures, different methods have been developed and explored \cite{Vogel2015b}. 
One unique way to tune the spontaneously formed structures is to assemble the colloidal particles (of sizes from $2$ nm to 1 $\mu$m) under spherical confinement induced by evaporating micro-emulsion droplets or larger, millimeter-sized dispersion droplets \cite{Marin2012, Wang2013, Vogel2015, Wintzheimer2018}.
Colloidal spheres typically crystallize into face centered cubic (FCC) crystal phases. When this type of crystallization occurs under spherical confinement, unique optical properties can arise, i.e., structural colors, in the final clusters. In these cases, cluster diameters can range from <50 $\mu$m \cite{Vogel2015col} to the millimeter-scale \cite{Rastogi2008,Rastogi2010}. Furthermore, control over the number of particles inside a dispersion droplet has been shown to produce novel structures, such as icosahedral clusters \cite{deNijs2015, Wang2021} and magic number clusters \cite{Wang2018b}. Depending on the initial colloidal volume percent and the drying substrate properties, dispersion droplets drying on superhydrophobic surfaces can form various cluster shapes, e.g., doughnuts or flattened discs \cite{Rastogi2010, Pauchard2004} and spheres \cite{Rastogi2008,Marin2012}. 

Besides finely tuning the assembly process, improvements in the synthesis of anisotropic colloidal particles have opened the path for more structurally varied assemblies \cite{Hueckel2021}. Consequently, there is a great diversity within the  colloidal assemblies of different anisotropic and polyhedral shapes \cite{Damasceno2012,VanDamme2020,Yuan2019}.
Combining spherical confinement with anisotropic shapes can lead to even more complex micro structures.  For instance, simulations of different polyhedral shapes showed that spherical confinement leads to unique clusters distinctly different from the packings of spheres \cite{Teich2016}. For rod-like colloids, it was found that their disordered so-called supraparticles possess unique scattering enhancing properties in contrast to their spherical counterparts \cite{Jacucci2021}.

Among available anisotropic particles, the superball shape, which encompasses the transition from a sphere to a cube via a rounded cube, has become of interest due to its unique shape-dependent phase behaviour \cite{Jiao2009,Batten2010,Ni2012} and experimental availability \cite{Zhang2011,Brunner2017, Rossi2011}.
For hard superballs,  the densest packing is a rhombohedral-like crystal structure that, with increasing asphericity of the superballs, continuously evolves from an FCC lattice (spheres) to a simple cubic (SC) lattice (perfect cubes) \cite{Jiao2009, Zhang2011,Brunner2017, Meijer2017,Batten2010}. Before assembling into their densest packings from a liquid phase, superballs can form a plastic crystal (or rotator) phase which has translational order and rotational mobility \cite{Ni2012, Meijer2017}. Experimentally, the phase diagram of micron-sized superballs has been studied via crystallization under gravitational settling in a capillary \cite{Meijer2017} and depletion-assisted assembly \cite{Rossi2011, Rossi2015}. For nano-sized superballs, the dense packings have also been investigated \cite{Zhang2011,Brunner2017}. These studies also confirmed the formation of dense rhombohedral-like structures of which the exact packing is dependent on the aphericity of the superball shape.

In addition, the self-assembly of micron-sized superballs into monolayers driven by solvent evaporation processed on flat substrates was investigated \cite{Meijer2011, Meijer2019}. These studies showed that solvent flow and immersion capillary forces, which can reach $10^6$ kT for micron-sized colloids, will lead to the formation of similar dense rhombohedral-like crystal structures but also revealed that the out-of-equilibrium process induces the formation of lower density crystal lattices with hexagonal-like structures as well as disordered packings.
%
%Furthermore, a change in the interparticle interaction between the superballs will further change the order in the self-assembled structures.
%
%For example, with depletion-assisted assembly, superballs can form SC crystals, \cite{Rossi2011} and by changing the ratio between the depletant and the superball sizes, solid-solid crystal transitions were found to occur \cite{Rossi2015}.
%
%In addition, monolayers of superballs, formed via solvent evaporation, form  both expected densest packings with additional switching between these \cite{Meijer2019}, while a mechanical unidirectional rubbing method results in alignment of the particle faces and sliding phases \cite{TenNapel2021}. 

So far, only a few studies have investigated the effect of spherical confinement on the assembly of superball (or cubic) particles and the relationship between particle asphericity and the resulting densely packed structure symmetries. 
For superball nanoparticles with increasing asphericity, the confined structure changes from icosahedral clusters with strong local orientation to simple-cubic structures \cite{Wang2018}. Cubic nanoparticles, with very sharp edges, assemble at the surface and in bulk  of the droplet and form densely packed clusters with random orientations throughout the confined structure  \cite{Agthe2016}. Moreover, it has been shown that when nanocubes form dense supraparticles, the finite corner roundness and the surface tension of the confining droplet determine the superstructure morphology \cite{Tang2020}. 
To understand the final self-assembled supraparticles' structure formed by superballs, it is crucial to understand the dynamics of the particles during droplet evaporation. For this, small angle x-ray scattering (SAXS) is a very powerful tool that allows visualization of the assembly process of different nanoparticles during spherical confinement in detail \cite{Sen2014,Agthe2016,Marino2018,Montanarella2018} and has been used to resolve micron-sized superball particle assembly \cite{Meijer2017}.

In this work, we use in-situ synchrotron small angle X-ray scattering  with microradian resolution ($\mu$rad-SAXS) to reveal the crystallization process of micron-sized silica superball shells under spherical confinement through droplet drying and elucidate the structural transitions during assembly. We are able to follow the droplet evaporation process in detail and identify different stages during the process. As the solvent evaporates, the particle concentration increases. After this stage, dewetting of the superball shells occurs before full solvent evaporation. We find that crystalline structures start to form as particle concentration increases, and that the structures remain throughout the later stages. In addition, we find that as we increase the asphericity of the superballs, the assembled supraparticles' structures are polycrystalline assemblies of their predicted rhombohedral-like lattices.  

%%%%%%%%%%%%
\begin{figure*}
    \centering
    \includegraphics[width=1\textwidth]{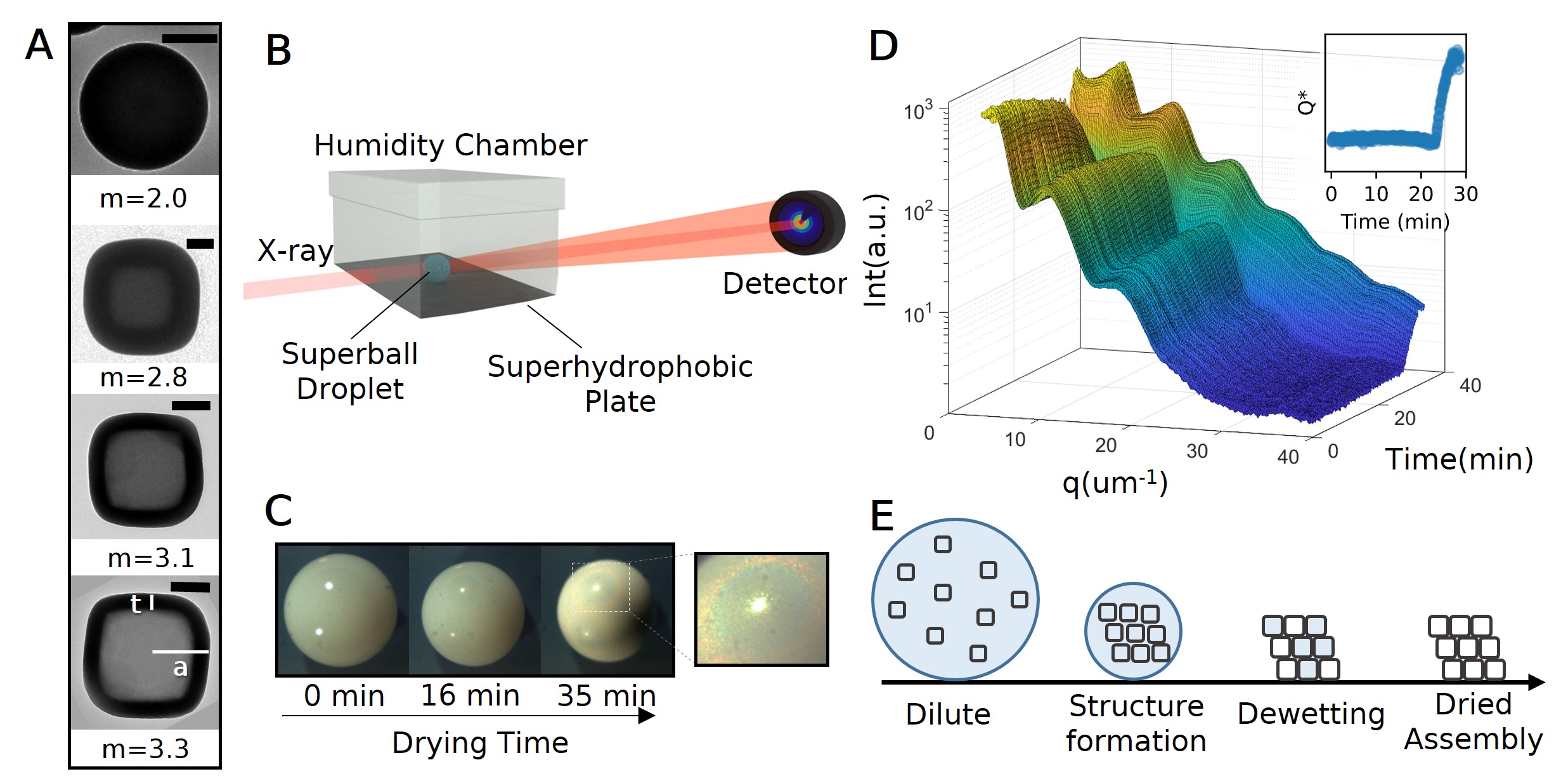}
    \caption{(A) Transmission electron microscopy images of a solid silica sphere and hollow silica superballs with different shape parameters, $m$, shell thicknesses, $t$, and radii, $a$. Scale bars are $200$ nm. (B) Schematic representation of the in-situ SAXS setup. A dispersion droplet on a superhydrophobic surface is placed inside a humidity chamber. Scattered x-rays are collected on a 2D detector with a wedge-shaped beam-stop. Schematic is not to scale.
    (C) A time sequence of an evaporating droplet of 3 $\mu$L containing superballs with $m=3.1$ collected with a digital camera. The droplets have a radius of approximately 1-2 mm. (D) Time-resolved radially-averaged SAXS profiles of the drying process of a droplet containing superballs with $m=2.6$ at 15vol\%. The inset shows the normalized scattering power $Q^*$ during the drying process. (E) A schematic representation of the different observed stages of drying: Dilute (non-interacting), structure formation, dewetting, and the final dried assembly.}
    \label{fig:Exp}
\end{figure*}
\section*{Results}
To study the self-assembly process of superball particles, we synthesized four samples of hollow silica superballs. We note that superballs are a family of shapes defined by $\left|\frac{x}{a}\right|^m +\left|\frac{y}{a}\right|^m+\left|\frac{z}{a}\right|^m \geq 1$ where $x$, $y$ and $z$ are the unit axis pointing toward the superball faces, $a$ is $\frac{1}{2}$ the distance between two opposite faces, and $m$ is the shape parameter\cite{Jiao2009, Elkies1991}. To create superballs with a range of m values, we coat hematite superball seeds in silica where increasing the silica shell thickness, $t$, decreases the resulting $m$ values \cite{Rossi2015}. The hematite seeds are etched away through the porous silica superball shell (Fig \ref{fig:Exp}A). For example, $m=2.6$ particles have shells of $t=120$ nm and particle sizes of $2\cdot a=1240$ nm with a size dispersity of 4\%. A table of all sample parameters is provided in supplementary Table S1.  For particles with $m=2$ (spheres), we grew two sets of silica particles without a template which resulted in solid silica spheres with radii of $r=206$ nm and $r=233$ nm \cite{Chen1998}. 

We fabricate assemblies of colloidal superballs by drying $1.5$ $\mu$L dispersion droplets of 15 vol\% superballs in MiliQ water on the surface of a superhydrophobic substrate. We add small pinning positions to the surface of the substrate with a needle tip to prevent droplet rolling. These types of dispersion droplets dry via diffusion of the water vapor into the air \cite{Pauchard2004}.

To investigate how the structure assembles under spherical confinement, we performed in-situ SAXS measurements on a series of drying dispersion droplets where each contained different $m$ valued superballs, as schematically shown in Fig \ref{fig:Exp}B.  To control the evaporation rate throughout experiments, droplets were dried at $\sim$64\% relative humidity in a custom-made chamber (Supplementary Fig S1). 
As water evaporates, the droplet contracts from the initial diameter of $1.5-2$ mm to $\sim 1$ mm (Supplementary Fig S2) and eventually yields a closed-packed assembly which is apparent from the rise in iridescent colors resulting from the Bragg scattering of light as seen in Fig \ref{fig:Exp}C \cite{Vogel2015, Rastogi2008}. 
The synchrotron X-ray beam, of width $500$ $\mu$m, passed through approximately the center of each droplet, and 2D images were recorded every $\sim 4$ seconds by the high-resolution 2D X-ray detector for 30-35 minutes as the droplet dried. From the 2D images, the 1D radially averaged scattering profiles were extracted. A typical set of scattering profiles collected in time during the droplet drying process for superballs, with $m=2.6$, is displayed in Fig \ref{fig:Exp}D. Time resolved scattering profiles for all investigated droplets are included in Supplementary Figs S3,S4.

%
%%%%%%%%%%%%
% Figure 2 %
%%%%%%%%%%%%
%
\begin{figure}
    \centering
    \includegraphics[width=0.4\textwidth]{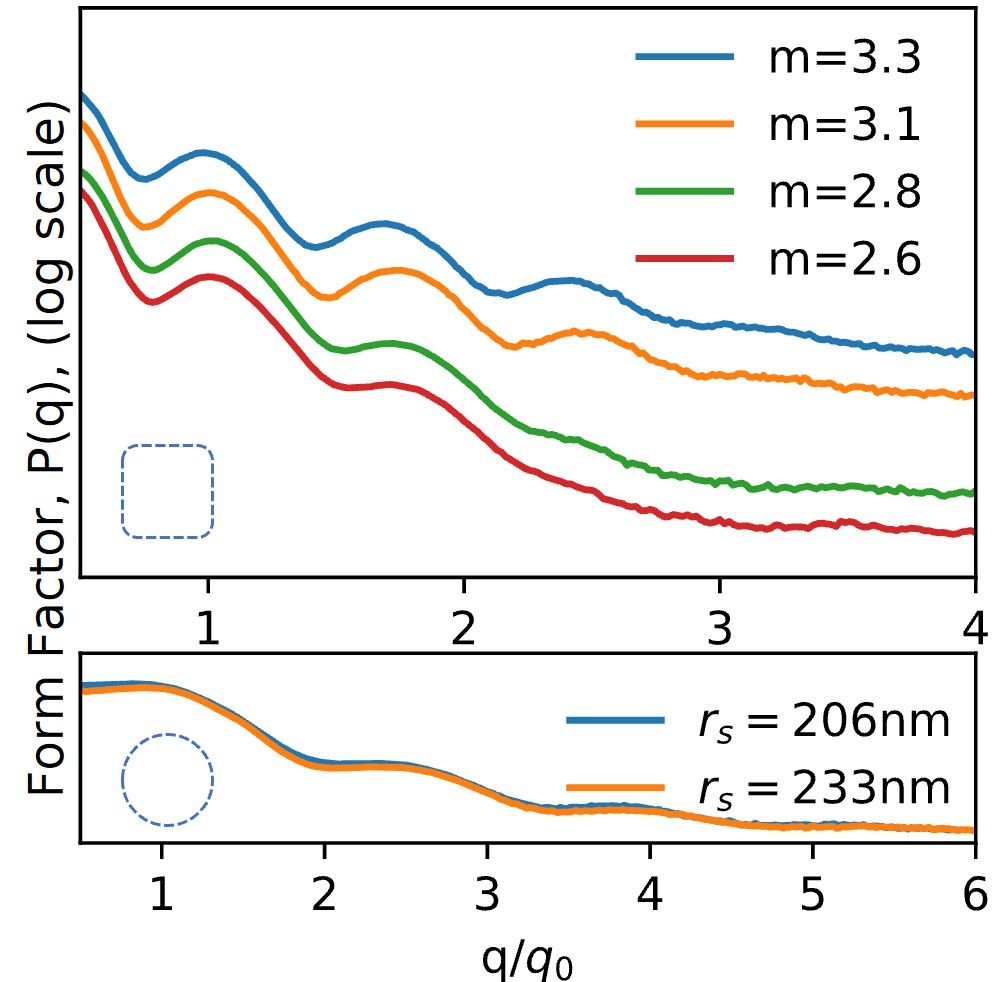}
    \caption{Extracted form factor for hollow silica superball and solid silica sphere experiments. The difference between sphere experiments is the particle radius where $r_A=233$ nm and $r_B=206$ nm. }
    \label{fig:FF}
\end{figure}

Generally, the scattering intensity, $I(q)$, from a collection of discrete particles is dependent on the form factor, $P(q)$, and the structure factor, $S(q)$. $P(q)$ arises from the particles' shape and size, while $S(q)$ arises from the particles' organization.
The total scattering intensity, $I(q)$, for superball particles can be approximated by the $I(q)$ for monodisperse spherical particles \cite{Meijer2017, Dekker2020b}. In this approximation,  $I(q)\propto \phi_{part}\cdot P(q) \cdot S(q)$ where $\phi_{part}$ is the particle volume fraction \cite{Agthe2016}. 
In addition, the overall droplet drying process can be monitored via the normalized scattering power, $Q^*$, which is defined by $Q^*=Q/Q_{max} \propto \Delta \rho \cdot \phi_{SB}(1-\phi_{SB})$ where $Q=\int^{\infty}_0 q^2 \cdot I(q) dq$ is the Porod invariant, $Q_{max}$ is the maximum $Q$ value, $\phi_{SB}$ is the volume fraction of superballs, and $\Delta \rho$ is the scattering length density (SLD) which arises from the difference in electron density between the particle and surrounding medium \cite{Sen2014}. 

Throughout the droplet drying process, we observe four distinct regimes in the $I(q)$ patterns as shown schematically in Fig \ref{fig:Exp}E.
Initially, the superball particles in the droplet are in a dilute, non-interacting, fluid state where we can assume that $S(q) \approx 1$ and that $P(q)$ dominates the scattering profile. We note that we assume $S(q) \approx 1$  in the observed experimental $q$ region due to the lack of additional long range effects from the superballs\cite{Zheng2011} because we consider the superballs to be hard particles. We check this assumption via the Percus-Yevick $S(q)$ for a $\phi_{SB}=15$ volume \% (Supplementary Fig S5). During the first drying period, the evaporation and confinement do not influence the system, which is evident from the limited changes in the patterns and the little variation in $Q^*$. 
As water evaporates from the droplet, the confining volume decreases, and the concentration of particles increases. As a result, interparticle interactions arise, leading to $S(q) \neq 1$, and $S(q)$ peaks start to appear in the $I(q)$ profiles indicating structure formation.
Near the final stage of drying, dewetting occurs where the remaining water inside and outside the particles evaporates, and the surrounding medium of the particles becomes air. This change of medium from water to air dramatically increases the difference between SLDs of the colloids and the surrounding medium, $\Delta \rho$.
This step in the drying process leads to a change in the overall intensity of the 1D profiles and appears as a rapid increase in $Q^*$. Therefore, we utilize the inflection point of $Q^*$ to determine the beginning of dewetting (see inset in Fig \ref{fig:Exp}D).  
For all investigated superballs and their droplets, dewetting began at similar times (at 24.1 minutes for $m=2.6$, at 21.7 minutes for $m=2.8$, $3.1$, and $3.3$, and at 22.9 minutes for both $m=2$ containing samples). Discrepancies between dewetting times are likely due to the $2-5$ minute time variation in placing the droplet and beginning the SAXS measurements.  
Eventually, all of the remaining water in the system evaporates and only the dried assembly of particles remains. When the dried assembly is formed, the 1D profiles and $Q^*$ plateau. During the SAXS measurements, not all investigated droplets formed a dried assembly (Supplementary Fig S3,S4).

Before characterizing the full drying process, we first characterize the particles in the initial, dilute state where $P(q)$ dominates the scattering profile. 
$P(q)$ was determined to be the initial free particle scattering profiles when $t=0$, $I_{t=0}(q)$ and is shown for all investigated particle shapes with $2\leq m\leq 3.3$ in Fig \ref{fig:FF}. 
As is well known for hollow spherical shells, the P(q) of a hollow object will be dependent on the details of the shell thickness and the contrast between the solvent, shell, and core \cite{Pedersen1997}. In the case of hollow superballs, the P(q) is determined by the superball shell thicknesses ($t$), inner shape ($m_{inner}$), and
outer shape ($m_{outer}$) parameters \cite{Meijer2017}. The slight variations visible between the $P(q)$ of the different hollow superballs seen in Fig \ref{fig:FF} can be attributed due to these differing parameters.
In addition, it is clear that the $P(q)$ minima are not very distinct, which we attributed to a small size dispersity ($3-5\%$) of the particles \cite{Rieker1999}.
It is important to note, for drying hollow superballs, using $I_{t=0}(q)$ as $P(q)$ is only valid when all the particles' cores match the surrounding medium's electron density. When dewetting occurs, particle cores do not empty at the same time and $P(q) \neq I_{t=0}(q)$ as shown schematically in Fig \ref{fig:Exp}E. Additionally, $S(q)$ changes as neighboring particles no longer posses the same SLD (for more details see Supplementary Fig S6).   

%
%%%%%%%%%%%%%
%  Figure 3 %
%%%%%%%%%%%%%
\begin{figure}[t]
    \centering
    \includegraphics[width=0.4\textwidth]{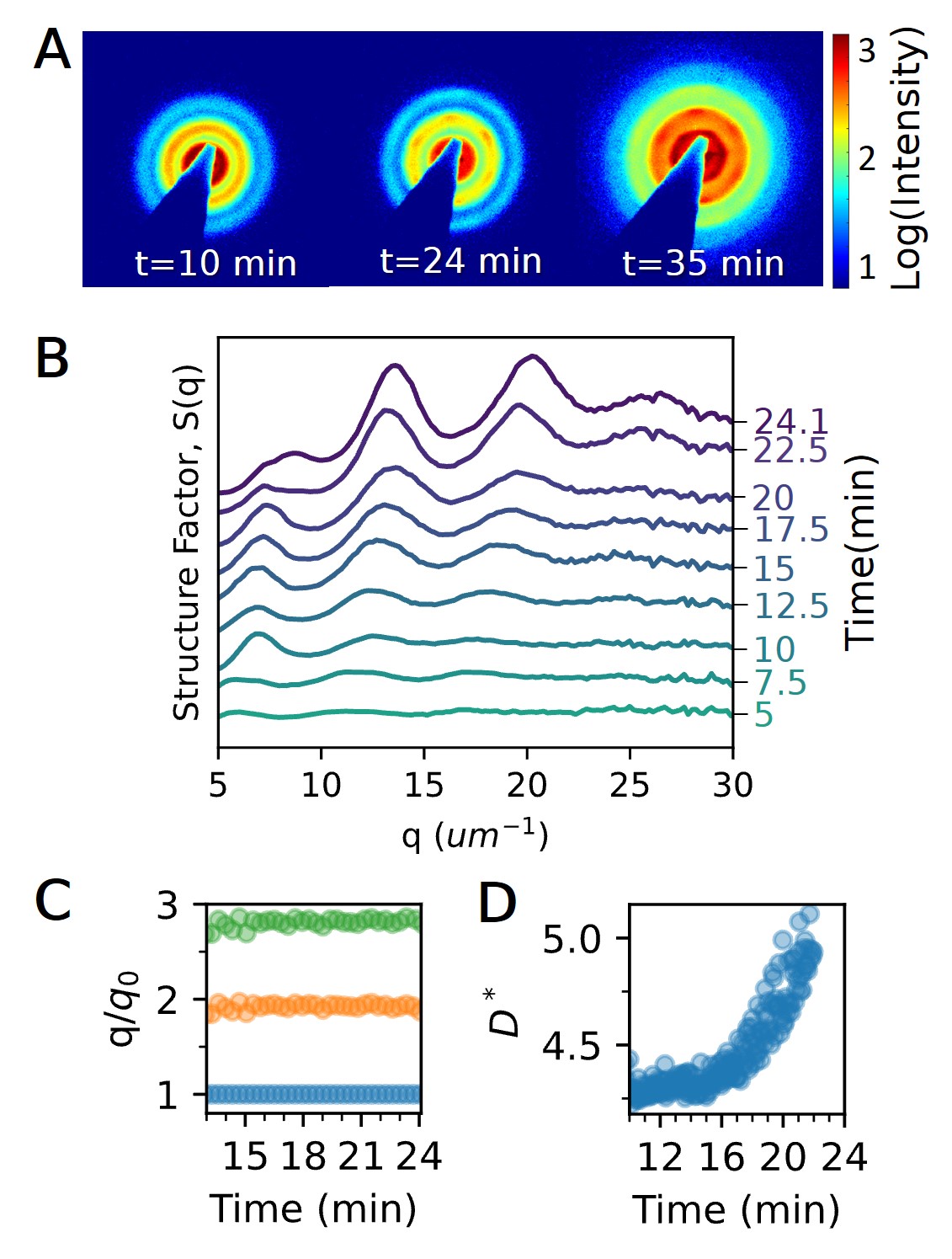}
    \caption{Structural evolution for a drying droplet containing superballs with $m=2.6$. (A) Selected 2D SAXS patterns showing the rise of Bragg peaks during drying. (B) Selected $S(q)$ curves extracted until the dewetting phase. (C) The ratio of peak positions to the first peak location ($q/q_0$) at selected times.  (D) The average crystallite size normalized to the particle size ($2 \cdot a$) over droplet drying.}
    \label{fig:SFPeak}
\end{figure}
We can visualize the drying process by examining the 2D scattering images as done in Fig \ref{fig:SFPeak}A for a drying droplet with $m=2.6$ superballs. Here, we observe isotropic scattering profiles from the fluid until Bragg peaks appear in the first ring after 10 minutes, indicating the formation of crystalline structures. These $S(q)$ peaks are present through the dewetting point ($t=24.1$ minutes). Compared to distinct single crystalline 2D SAXS patterns observed for 3D superball sediments \cite{Meijer2013, Meijer2017}, the radial width of these peaks is larger and the fact that higher order peaks are absent indicate that the resultant structure is polycrystalline where the domains have only weak orientational correlations with each other. For all investigated superball dispersion droplets, we find similar scattering patterns. Although as $m$ increases, the peaks broaden, and for $m=3.3$ only a ring can be seen (see Supplementary Fig S8).

To further understand the structure formation, we examine the 1D $S(q)$ curves. We can extract $S(q)$ by dividing $I(q)$ with the $P(q) \approx$ $I_{t=0}(q)$ \cite{Zhang2011} up to dewetting phase where the patterns start to change dramatically (as discussed above).  
Fig \ref{fig:SFPeak}B shows selected $S(q)$ curves for superballs with $m=2.6$ up to $t=24.1$ min. 
It can be seen that three consistent, albeit broad, peaks develop in the curves around $t=10$ min indicating that structure formation has begun. During drying, all peak positions continuously shift towards higher $q$ values demonstrating decreasing inter-particle distances between the superball particles (Supplementary Fig S7). After $t=20$ min, a shoulder emerges from the $q_0$ peak.

To reveal how the structure evolves, we follow the ratio between the peak positions and the first peak, $q/q_0$ \cite{Zhang2011,Meijer2017}. Peak positions and widths were determined by fitting the peaks with a Voigt function\cite{Olivero1977}. Fig \ref{fig:SFPeak}C shows the peak position ratios over time for the $m=2.6$ superball droplet. The ratio remains relatively constant throughout droplet drying and suggests that no significant phase transitions occur. 
%We find the same trend for all experimental droplets ( see Fig S2,S3). 
%
From the $q_0$ peak width, the average crystallite size can be calculated using the Scherrer equation \cite{Patterson1939}.
To easily  compare the result between different cubes, the average crystallite size, $D$, is normalized by dividing the superball side length, $(2 \cdot a)$,  to find $D^*=\frac{D}{2\cdot a}$. Fig \ref{fig:SFPeak}D shows that $D^*$ begins to increase rapidly after 16 minutes as the droplet dries. This indicates that the average crystalline domain grows. 
%
%
%%%%%%%%%%%%%%
%  Figure 4  %
%%%%%%%%%%%%%%
%
\begin{figure*}[t]
    \centering
    \includegraphics[width=0.9\textwidth]{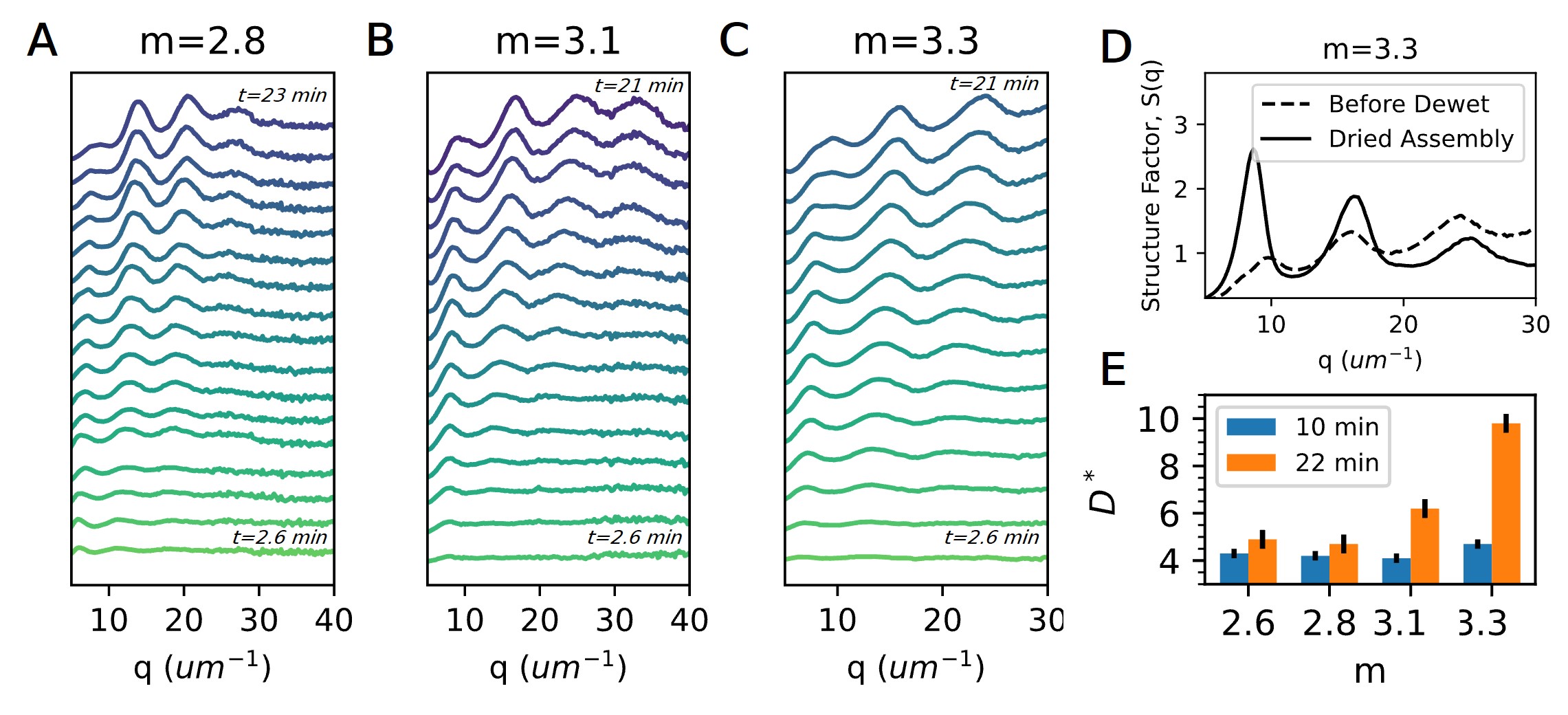}
    \caption{Structural evolution of superball dispersion droplets during drying. Selected $S(q)$ curves extracted until the dewetting phase for (A) $m=2.8$, (B) $m=3.1$, and (C) $m=3.3$ with a time-step of $1.2$ min between curves. (D) Comparison of $S(q)$ curves from before dewetting (black, dashed) to the final dried assembly (black, solid) for $m=3.3$. (E) The normalized average crystallite size for all superball systems at 10 minutes (blue bars) and 22 minutes (orange bars).}
    \label{fig:SF}
\end{figure*}

To understand the influence of particle shape during assembly, we explore how $S(q)$ evolves for all dispersion droplets of superballs with $2.6\leq m \leq 3.3$. 
Figs \ref{fig:SF}A-C show the temporal $S(q)$ curves for $m=2.8$, $m=3.1$, and $m=3.3$. In all droplets, we observe a trend similar to $m=2.6$ where $S(q)>1$ at $t \approx 7$ minutes. The peak positions shift to higher $q$ values, and a shoulder emerges from the $q_0$ peak. 
To confirm that the structure remains after dewetting occurs, we can extract $S(q)$ of a dried assembly and compare to the structure before dewetting by rescaling $P(q)$ \cite{Pedersen1997}. Only particles with $m=3.3$ formed a completely dried assembly which was determined by a plateau at the end of $Q^*$ (Supplementary Figs S3-S4). 
The solid black curve in Fig \ref{fig:SF}D depicts the structure of the dried assembly and is compared to $S(q)$ before dewetting (dashed black curve). The increase in intensity is due to the change in surrounding medium \cite{Pedersen1997}. Otherwise, there are no significant changes to $S(q)$ after dewetting.
Furthermore, we compare the average crystallite size, $D^*$, over time for all investigated droplets in Fig \ref{fig:SF}E. As the droplets dry, we observe that the average crystallite size increases for all investigated droplets. Indeed, the average crystallite size at the onset of structure formation ($\approx 10$ minutes) spans approximately 5 particle diameters for all droplets. At 22 minutes before dewetting occurs, $D^* \approx 5$ particles per crystallite for $m=2.6$ and $m=2.8$. Then, $D^*$ increases with the $m$ value where the most cubic-like particle assembly ($m=3.3$) has $D^* \approx 9.8$ particles per crystallite.

\begin{figure*}[t]
    \centering
    \includegraphics[width=0.95\textwidth]{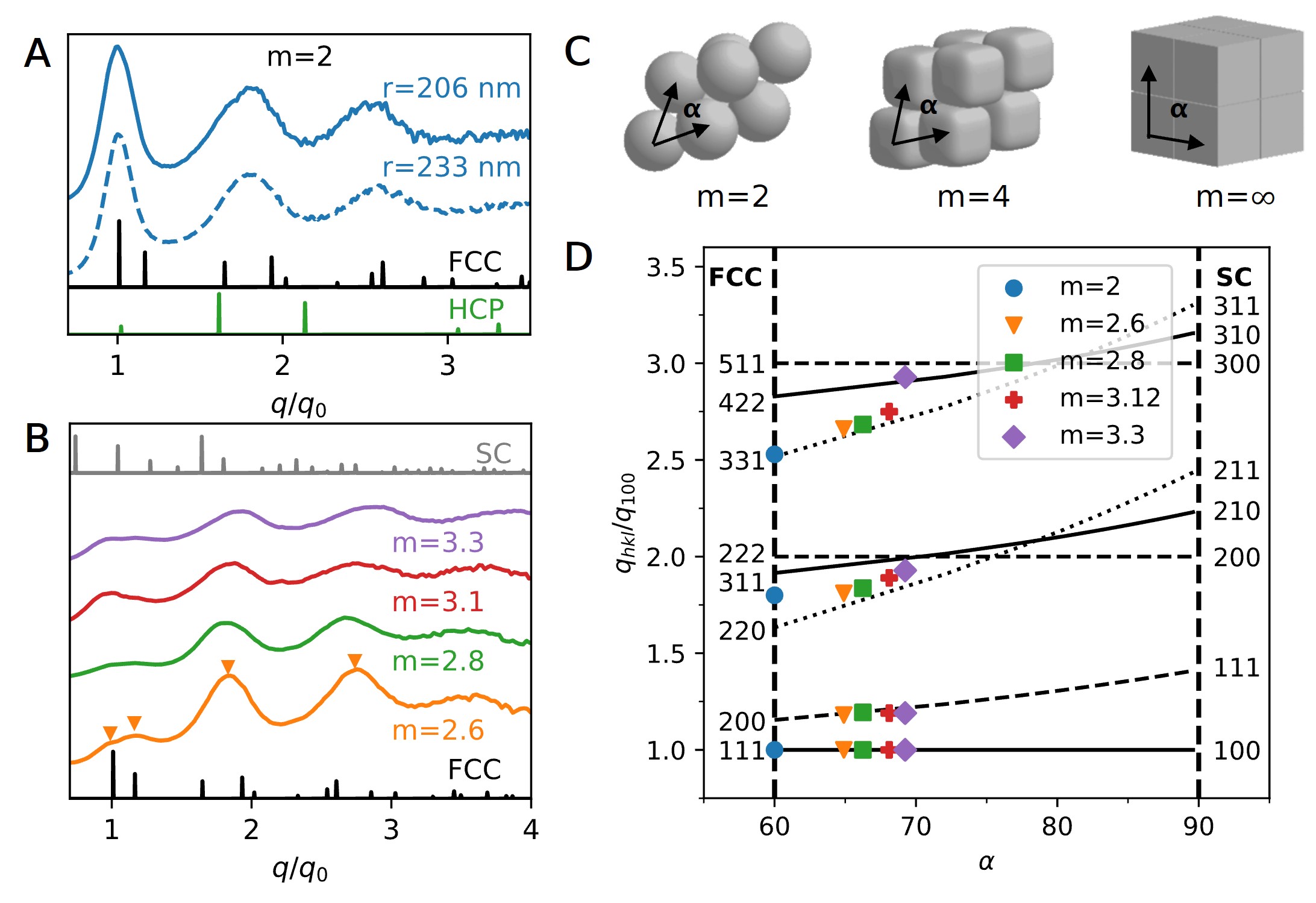}
    \caption{Evolution of the structure over $m$. (A) Stacked plot of extracted structure factors before dewetting for all sphere experiments with model hexagonal closed packed (HCP) and face-centered cubic (FCC) lines. (B) Stacked plot of extracted structure factors before dewetting for all superball experiments with model simple-cubic (SC) and face-centered cubic (FCC) lines. (C) Schematic of the angle dependent rhombic structure as the $m$ value transitions from $2\geq m \geq \infty$. (D) Model lines are plotted of the allowed ratio of the $(hkl)$ diffraction peak position over the angle, $\alpha$. Experimental values (m=2, 2.6, 2.8, 3.1, and 3.3) of the peak position ratios are overlaid on top (markers). }
    \label{fig:hkl}
\end{figure*}

To determine the crystal structures that have formed in the droplets, we first extract $S(q)$ for the spherical particle dispersion droplets just before dewetting (Fig \ref{fig:hkl}A). To compare different dispersions, we re-scaled $q$ of all the $S(q)$ curves by the corresponding $q_0$ position of the first $S(q)$ peak. In general, spherical colloids do not assemble into pure FCC lattices, but they form a random hexagonal close packed (RCHP) structure that is a mixture of both the FCC and hexagonal close packed (HCP) lattices \cite{Petukhov2003}. Therefore, we compare the $S(q)$ curves from two separate droplets with different solid silica sphere particle radii ($R=206,233$ nm) to the expected $S(q)$ peaks for FCC (black model lines) and HCP (green lines) in Fig \ref{fig:hkl}A. We find that the spherical assemblies produced similar broad peaks near expected FCC peak positions for the $hkl$ planes $111$, $311$, $331$, and $440$.

On the other hand, as the $m$ value increases, we expect that perfect cubic particles pack into simple cubic (SC) lattices \cite{Agthe2016}. Therefore, in Fig \ref{fig:hkl}B, we compare the curves to the expected $S(q)$ peaks for FCC (black model lines) and SC lattices (gray model lines). For all $m$, it is clear that the peaks do not perfectly match either the FCC peaks or the SC peaks. Instead as we increase $m$, we observe a slight shift of peak positions away from $q_0$. 
%
%%%%%%%%%%%%%%%%
%  Figure 5    %
%%%%%%%%%%%%%%%%
%
Indeed, simulations and experiments have shown that the equilibrium structure formed by superballs, with $m \geq 2.30$, is a densely packed rhombic crystal lattice, described by the C$_{1}$-lattice \cite{Jiao2009,Batten2010,Ni2012,Meijer2017}.
The C$_{1}$-lattice is dependent on the superball $m$ parameter and describes a continuous transition from an FCC structure for  $m=2$ to an SC structure for $m \rightarrow \infty$ as shown in Fig \ref{fig:hkl}C.

To compare our assembled structures to the C$_1$-lattice, we calculate how the $S(q)$ peaks evolve as the $m$ values increases. The lattice vectors for the C$_1$-lattice are defined by Jiao \textit{et al.} \cite{Jiao2009} as $\mathbf{e_1} = -2(s+2^{-1/m})\mathbf{i} + 2s\mathbf{j}+2s\mathbf{k}$, $\mathbf{e_2} = -2s\mathbf{i} +2s\mathbf{j} +2(s+2^{-1/m})\mathbf{k}$, and $\mathbf{e_3} = -2s\mathbf{i}+2(s+2^{-1/m})\mathbf{j} + 2s\mathbf{k}$
where $s$ is the smallest positive root of the equation $(s+2^{-1/m})^m + 2s^m-1=0$. The structure varies as $m$ increases by changing the angle, $\alpha$, between lattice vectors: $\alpha=\arccos\frac{\mathbf{e_i \cdot e_j}}{\mathbf{|e_i||e_j|}}$ where $i,j=1,2,3 (i\neq j)$. It has been shown that when superballs assemble into their densest packing, we can directly map their expected angle, $\alpha$, from the experimental $m$ value \cite{Meijer2017}. We note that as $m$ increases from 2 to 3.3, the angle, $\alpha$, increases from $60^{\circ}$ to $69^{\circ}$ (Supplementary Fig S11). Achieving higher angles ($\alpha > 80^{\circ}$) requires particles with high $m$ values ($m>10$). 

Then, we can follow the transition between FCC and SC lattices of the C$_1$ lattice via the ratio of the diffraction peak position for $60^{\circ}< \alpha < 90^{\circ}$. This ratio is defined as 
\begin{equation}
    \frac{q_{hkl}}{q_{100}} = \big[\frac{(h^2+k^2+l^2)\sin{\alpha}^2}{\sin{\alpha}^2}+\frac{2(hk+kl+lh)(\cos{\alpha}^2-\cos{\alpha})}{\sin{\alpha}^2}\big]^\frac{1}{2}
\end{equation}
%\begin{multline}
%    \frac{q_{hkl}}{q_{100}} = \big[\frac{(h^2+k^2+l^2)\sin{\alpha}^2}{\sin{\alpha}^2}\\
%    +\frac{2(hk+kl+lh)(\cos{\alpha}^2-\cos{\alpha})}{\sin{\alpha}^2}\big]^\frac{1}{2}
%\end{multline}
where $q_{hkl}$ are the peak positions of allowed $(hkl)$ reflections.
Fig \ref{fig:hkl}D shows how the ratios of several $hkl$ peak positions are expected evolve according to eq.(1) for angles from $60\leq \alpha \leq 90$ (lines). The respective FCC and SC $hkl$ values related to the calculated rhombohedral planes are included. Additionally, we overlay the experimental data points (symbols) on top of the lines for comparison.
As $\alpha$ increases, the diffraction ratio generally grows for most $hkl$ planes. Higher $q$ value peaks tend to increase more rapidly compared to lower $q$ value peaks. However, that is not the case for all $hkl$ planes such as the horizontal lines associated to the transitions into the SC $h00$ planes.
While the experimental values are only available up to $m=3.3$, corresponding to $\alpha \approx 70$, the peak ratio appears to generally follow the predicted trends. We observe that the emerging peak off of $q_0$ closely follows the FCC $200\rightarrow110$ SC plane transition. Similarly, the higher-order peaks also increase with increasing $m$.
%
%
%%%%%%%%%%%%%%%
% Discussion  %
%%%%%%%%%%%%%%%
%
%
\section*{Discussion}
Our $\mu$rad-SAXS results show that the spherical confinement, due to drying dispersion droplets of different superball colloids, leads to the formation of polycrystalline assemblies with their expected rhombohedral C$_1$-lattice packing \cite{Jiao2009}. We further find that the average crystalline domain increases with the particles' $m$ value. Our findings are in part in agreement with the results of Wang \textit{et al.}\cite{Wang2018} who showed that for small droplets, with $\approx$2000 cubic nanoparticles, there are short range correlations between the particle positions of dried superball assemblies with low $m$ values, and long-range correlation increases with increasing $m$. However, in contrast to their study, we do not find the formation of icosahedral clusters for low $m$ as the droplets investigated here are much larger and contain $\approx 2\times 10^9$ superball particles. In this respect, our results are more in line with studies focusing on droplets with a larger number of particles. For nanocubes with $m>5$, Agthe \textit{et al.}\cite{Agthe2016} showed that spherical confinement in a levitating drop leads to polycrystalline domains where domains at the surface were aligned to the droplet interface. For micron-sized spheres, Marin \textit{et al.}\cite{Marin2012} showed that droplets on superhydropic surfaces, at high enough particle concentration, will form densely packed structures with many different crystalline domains also aligned to the droplet surface. The reason for this polycrystalline structure is that densification of a droplet leads to multiple nucleation sites when the droplet contains a large number of particles. 
Interestingly, we did observe the appearance of distinct six-fold Bragg peaks on top of the S(q) ring in the 2D $\mu$rad-SAXS patterns of dispersion droplets with $m<3.3$. This could indicate that while a large part of the surface crystallites are randomly oriented, the center part of the droplet may contain a structure with long-range order,  but more investigations are needed to confirm this.  

By following the full drying process, we showed that the system densifies and the crystalline structure starts to form as the droplet evaporates. Interestingly, the forming crystal structures did not undergo any phase transitions, which might be expected based on a previous study by Meijer \textit{et al.}\cite{Meijer2017} who showed that superball colloids, upon slow sedimentation, will first form a rotator FCC crystal before transforming into their densest rhombic packing. Clearly, the current assembly process during spherical confinement occurs via different pathways. We further revealed that a dewetting stage occurs, during which the water evaporates from between and inside the hollow colloids, leading to the dramatic change in the particle $P(q)$. Since the water does not evaporate from all of the porous hollow colloids at one instant, we do not observe such a significant change for the solid spheres, and this has not been observed for other droplet drying studies of solid particles as well. Thus these hollow particles allowed us to obtain more insight into the local evaporation process as we tracked the changing intensity in $P(q)$ and the corresponding $I(q)$. In addition, strong capillary forces acting on the colloidal particles during dewetting of the droplet, which can reach $10^6$ kT, \cite{Kralchevsky1994} do not lead to rearrangement of the crystalline structures that have formed, as the S(q) before and after dewetting are the same.

In conclusion, we investigated the structural evolution of colloidal superball assemblies during spherical confinement induced by a drying dispersion droplet. The presented results show that the structure of the assembled superballs nucleate into the predicted C$_1$-lattice. As the droplet evaporates, we found that the structure does not undergo any additional phase transitions throughout drying. The resulting assembly contains several small crystallites that increase in size as the superballs become more cubic-like indicating a polycrystalline structure, and the assembly exhibits optical reflections on its surface.

As our understanding of superball assemblies continues to advance, we anticipate that insights into the relationships between the constituent anisotropic particles and resulting assembly's structural color and mechanical properties will arise. It has been shown that by switching from spherical to rod-like particles, dramatic changes in the assembly's structural color can be obtained after spherical confinement \cite{Jacucci2021}. Similarly,  particle size and interparticle bonds influence the mechanical properties of the assembly \cite{Wang2021b}. Looking forward, we anticipate that by retaining the magnetic hematite seed inside the silica superball \cite{Meijer2013}, we can explore how the additional magnetic dipole interactions influence the assembly under spherical confinement. 

%
%%%%%%%%%%%%%
% Methods   %
%%%%%%%%%%%%%
%
\section*{Methods}
%=========
% Particle Preparation
%=========
\subsection*{Particle Preparation}

Hematite cube template particles were prepared following a sol-gel procedure \cite{Sugimoto1992}. Silica shells were grown following Ref. \citenum{Graf2003}.
The hematite templates were dissolved through the addition of hydrochloric acid to produce hollow shells. 
The complete particle synthesis procedure to obtain different $m$ values can be found in Ref. \citenum{Rossi2015}.
Silica spheres were prepared using a variation on the St\"{o}ber method \cite{Stober1968} from Ref. \citenum{Chen1998} to create spheres with diameter of $D \approx 500$ nm.
The obtained particles were characterized by transmission electron microscopy (TEM - Philips TECNAI12/20). Details on the properties of all prepared particles are presented in Supplementary Table S1.
%=========
% Assembly Process
%=========
\subsection*{Assembly Process}
%=========
% Superhydrophobic substrate
%=========
\subsubsection*{Superhydrophobic Substrate}

Superhydrophobic substrates with contact angles above 160$^{\circ}$ were prepared following a modified procedure from Deng \textit{et al.} \cite{Deng2012}. Briefly, a thin layer of soot from the flame of a paraffin candle is deposited on a cleaned glass substrate. Then, the substrate is coated with a thin silica shell by enclosing it in a desiccator with two open containers of 2mL tetraethyl orthosilicate and 2mL NH$_3$ (29 wt\%) for 24 hr. 
The substrate is then calcinated in an Evenheat Rampmaster RM2 furnace by heating at 600$^{\circ}$C for 2 hr. 
After, 350 $\mu$L of 1\% trichlorododeylsilane 
%(TDS) 
in chloroform is dropcasted on the substrate and allowed to evaporate. 
Droplet pinning positions are created on the substrate via notches on the surface. 
%=========
%  of Colloids
%=========
\subsubsection*{Spherical Confinement of Colloids}

Self assembled macrostructures of colloidal spheres and superballs were formed by depositing and drying 2-5 $\mu$L droplets of colloidal dispersions from 5 to 30 vol\% on a superhydrophobic substrate at the pinning positions. 
These are necessary to prevent the droplets from moving and escaping the superhydrophobic surface.
Evaporation rates can be controlled via drying inside a humidity chamber (See Supplementary Fig S1).
%=========
% Small Angle X-ray scattering (SAXS) measurements
%=========
\subsubsection*{Small Angle X-ray scattering (SAXS) measurements}

$\mu$rad-SAXS measurements were conducted at the  beamline BM26B DUBBLE \cite{Bras2003} at the European Synchrotron Radiation Facility (ESRF) in Grenoble, France using a microradian setup \cite{Petukhov2015} employing compound refractive lensens \cite{Snigirev1996}. The incident X-ray beam  with a photon energy of 13 keV ($\lambda$ = $\SI{1.0332}{\angstrom}$) was focused on a Photonic Science CCD detector with $4,008 \times 2,671$ pixels and a pixel size of $22 \times 22$ $\mu m$ at a distance of 7.11 m. The detector was protected from the direct beam by a wedge-shaped beam-stop. Substrates with the drying droplets were placed horizontally in the X-ray beam. Measurements were performed during the full evaporation process at the position of the initial center of the drying droplet. \\

%=================
%=================
% Acknowledgments 
%=================
%=================

\acknowledgments
We thank the Netherlands Organization for Scientific Research (NWO) for the provided beam-time. NWO is also acknowledged for financial support of L.R. through a VENI grant(680-47-446) and J.M.M. (016.Veni.192.119). L.B. acknowledges the support by a Studienstiftung des Deutschen Volkes research grant. We are also grateful to the DUBBLE personnel of the European Synchrotron Radiation Facility in Grenoble for assistance with the small angle X-ray scattering experiments.

%============
%============
% References
%============
%============

\pagebreak


\begin{thebibliography}{10}

\bibitem{Manoharan2015}
Vinothan~N. Manoharan.
\newblock Colloidal matter: Packing, geometry, and entropy.
\newblock {\em Science} \textbf{2015}, 349(6251)

\bibitem{Zhao2020}
Tianheng~H. Zhao, Gianni Jacucci, Xi~Chen, Dong-Po Song, Silvia Vignolini, and
  Richard~M. Parker.
\newblock Angular-independent photonic pigments via the controlled
  micellization of amphiphilic bottlebrush block copolymers.
\newblock {\em Advanced Materials} \textbf{2020}, 32(47):2002681

\bibitem{Vogel2015b}
Nicolas Vogel, Markus Retsch, Charles-André Fustin, Aranzazu del Campo, and
  Ulrich Jonas.
\newblock Advances in colloidal assembly: The design of structure and hierarchy
  in two and three dimensions.
\newblock {\em Chemical Reviews} \textbf{2015} 115(13):6265-6311.

\bibitem{Marin2012}
{\'A}lvaro~G. Mar{\'\i}n, Hanneke Gelderblom, Arturo Susarrey-Arce, Arie van
  Houselt, Leon Lefferts, Johannes G.~E. Gardeniers, Detlef Lohse, and Jacco~H.
  Snoeijer.
\newblock Building microscopic soccer balls with evaporating colloidal fakir
  drops.
\newblock {\em Proceedings of the National Academy of Sciences} \textbf{2012},
  109(41):16455--16458

\bibitem{Wang2013}
Tie Wang, Derek LaMontagne, Jared Lynch, Jiaqi Zhuang, and Y.~Charles Cao.
\newblock Colloidal superparticles from nanoparticle assembly.
\newblock {\em Chem. Soc. Rev.} \textbf{2013}, 42:2804--2823

\bibitem{Vogel2015}
Nicolas Vogel, Stefanie Utech, Grant~T. England, Tanya Shirman, Katherine~R.
  Phillips, Natalie Koay, Ian~B. Burgess, Mathias Kolle, David~A. Weitz, and
  Joanna Aizenberg.
\newblock {Color from hierarchy: Diverse optical properties of micron-sized
  spherical colloidal assemblies}.
\newblock {\em Proceedings of the National Academy of Sciences} \textbf{{2015}},
  112(35):10845--10850

\bibitem{Wintzheimer2018}
Susanne Wintzheimer, Tim Granath, Maximilian Oppmann, Thomas Kister, Thibaut
  Thai, Tobias Kraus, Nicolas Vogel, and Karl Mandel.
\newblock Supraparticles: Functionality from uniform structural motifs.
\newblock {\em ACS Nano} \textbf{2018}, 12(6):5093--5120

\bibitem{Vogel2015col}
Nicolas Vogel, Stefanie Utech, Grant~T. England, Tanya Shirman, Katherine~R.
  Phillips, Natalie Koay, Ian~B. Burgess, Mathias Kolle, David~A. Weitz, and
  Joanna Aizenberg.
\newblock Color from hierarchy: Diverse optical properties of micron-sized
  spherical colloidal assemblies.
\newblock {\em Proceedings of the National Academy of Sciences} \textbf{2015},
  112(35):10845--10850

\bibitem{Rastogi2008}
V. Rastogi, S. Melle, O.~G. Calderón, A.~A. García, M. Marquez, and O.~D. Velev.
\newblock Synthesis of Light-Diffracting Assemblies from Microspheres and Nanoparticles in Droplets on a Superhydrophobic Surface
\newblock {\em Advanced Materials} \textbf{2008}, 20(22):4263--4268

\bibitem{Rastogi2010}
V. Rastogi, A.~A. García, M. Marquez, and O.~D. Velev.
\newblock Anisotropic Particle Synthesis Inside Droplet Templates on Superhydrophobic Surfaces.
\newblock {\em Macromolecular Rapid Communications} \textbf{2010}, 31(2):190--195


\bibitem{deNijs2015}
Bart de~Nijs, Simone Dussi, Frank Smallenburg, Johannes~D. Meeldijk, Dirk~J.
  Groenendijk, Laura Filion, Arnout Imhof, Alfons van Blaaderen, and Marjolein
  Dijkstra.
\newblock {Entropy-driven formation of large icosahedral colloidal clusters by
  spherical confinement}.
\newblock {\em {Nature Materials}} \textbf{{2015}}, {14}({1}):{56--60}

\bibitem{Wang2021}
Da~Wang, Tonnishtha Dasgupta, Ernest~B. van~der Wee, Daniele Zanaga, Thomas
  Altantzis, Yaoting Wu, Gabriele~M. Coli, Christopher~B. Murray, Sara Bals,
  Marjolein Dijkstra, and Alfons van Blaaderen.
\newblock {Binary icosahedral clusters of hard spheres in spherical
  confinement}.
\newblock {\em {Nature Physics}} \textbf{{2021}}, {17},{128--134}

\bibitem{Wang2018b}
Junwei Wang, Chrameh~Fru Mbah, Thomas Przybilla, Benjamin~Apeleo Zubiri,
  Erdmann Spiecker, Michael Engel, and Nicolas Vogel.
\newblock {Magic number colloidal clusters as minimum free energy structures}.
\newblock {\em {Nature Communication}} \textbf{{2018}}, {9},{5259}

\bibitem{Pauchard2004}
L. Pauchard and Y. Couder.
\newblock Invagination during the collapse of an inhomogeneous spheroidal shell
\newblock {\em {Europhysics Letters}} \textbf{2004}, 66(5):667--673


\bibitem{Hueckel2021}
Theodore Hueckel, Glen~M. Hocky, and Stefano Sacanna.
\newblock {Total synthesis of colloidal matter}.
\newblock {\em {Nature Reviews Materials}} \textbf{2021}

\bibitem{Damasceno2012}
Pablo~F. Damasceno, Michael Engel, and Sharon~C. Glotzer.
\newblock Predictive self-assembly of polyhedra into complex structures.
\newblock {\em Science} \textbf{2012}, 337(6093):453--457

\bibitem{VanDamme2020}
Robin van Damme, Gabriele~M. Coli, Ren{\'{e}} van Roij, and Marjolein Dijkstra.
\newblock {Classifying Crystals of Rounded Tetrahedra and Determining Their
  Order Parameters Using Dimensionality Reduction}.
\newblock {\em ACS Nano} \textbf{2020}, 14(11):15144--15153

\bibitem{Yuan2019}
Ye~Yuan, Lufeng Liu, Wei Deng, and Shuixiang Li.
\newblock {Random-packing properties of spheropolyhedra}.
\newblock {\em Powder Technology} \textbf{2019}, 351:186--194

\bibitem{Teich2016}
Erin~G. Teich, Greg {Van Anders}, Daphne Klotsa, Julia Dshemuchadse, and
  Sharon~C. Glotzer.
\newblock {Clusters of polyhedra in spherical confinement}.
\newblock {\em Proceedings of the National Academy of Sciences of the United
  States of America} \textbf{2016}, 113(6):E669--E678

\bibitem{Jacucci2021}
Gianni Jacucci, Brooke~W. Longbottom, Christopher~C. Parkins, Stefan A.~F. Bon,
  and Silvia Vignolini.
\newblock Anisotropic silica colloids for light scattering.
\newblock {\em J. Mater. Chem. C} \textbf{2021}, 9:2695--2700

\bibitem{Jiao2009}
Y~Jiao, F~H Stillinger, and S~Torquato.
\newblock {Optimal packings of superballs}.
\newblock {\em Physical Review E} \textbf{2009}, 79(4):041309

\bibitem{Batten2010}
Robert~D. Batten, Frank~H. Stillinger, and Salvatore Torquato.
\newblock Phase behavior of colloidal superballs: Shape interpolation from
  spheres to cubes.
\newblock {\em Phys. Rev. E} \textbf{2010}, 81:061105

\bibitem{Ni2012}
Ran Ni, Anjan~Prasad Gantapara, Joost {De Graaf}, Ren{\'{e}} {Van Roij}, and
  Marjolein Dijkstra.
\newblock {Phase diagram of colloidal hard superballs: From cubes via spheres
  to octahedra}.
\newblock {\em Soft Matter} \textbf{2012}, 8(34):8826--8834

\bibitem{Zhang2011}
Yugang Zhang, Fang Lu, Daniel {Van Der Lelie}, and Oleg Gang.
\newblock {Continuous phase transformation in nanocube assemblies}.
\newblock {\em Physical Review Letters} \textbf{2011}, 107(13):135701

\bibitem{Brunner2017}
J.~Brunner, I.~A. Baburin, S.~Sturm, K.~Kvashnina, A.~Rossberg, T.~Pietsch,
  S.~Andreev, E.~Sturm~(née Rosseeva), and H.~Cölfen.
\newblock Self-assembled magnetite mesocrystalline films: Toward structural
  evolution from 2d to 3d superlattices.
\newblock {\em Advanced Materials Interfaces} \textbf{2017}, 4(1):1600431


\bibitem{Meijer2017}
Janne~Mieke Meijer, Antara Pal, Samia Ouhajji, Henk~N.W. Lekkerkerker,
  Albert~P. Philipse, and Andrei~V. Petukhov.
\newblock {Observation of solid-solid transitions in 3D crystals of colloidal
  superballs}.
\newblock {\em Nature Communications} \textbf{2017}, 8:14352

\bibitem{Rossi2011}
Laura Rossi, Stefano Sacanna, William~T.M. Irvine, Paul~M. Chaikin, David~J.
  Pine, and Albert~P. Philipse.
\newblock {Cubic crystals from cubic colloids}.
\newblock {\em Soft Matter} \textbf{2011}, 7(9):4139--4142

\bibitem{Rossi2015} %FIX
Laura Rossi, Vishal Soni, Douglas~J. Ashton, David~J. Pine, Albert~P. Philipse, Paul~M. Chaikin, Marjolein Dijkstra, Stefano Sacanna, and William~T.M. Irvine
{Shape-sensitive crystallization in colloidal superball fluids}.
\newblock {\em Proceedings of the National Academy of Sciences of the United
  States of America} \textbf{2015}, 112(17):5286--5290
\bibitem{Meijer2011}
Janne-Mieke Meijer, Fabian Hagemans, Laura Rossi, Dmytro~V. Byelov, Sonja~I.R. Castillo, Anatoly Snigirev, Irina Snigireva, Albert~P. Philipse, and Andrei~V. Petukhov
\newblock Self-Assembly of Colloidal Cubes via Vertical Deposition
\newblock{\em Langmuir} \textbf{2012}, 28(20):7631--7638

\bibitem{Meijer2019}
Janne-Mieke Meijer, Vera Meester, Fabian Hagemans, H.N.W. Lekkerkerker,
  Albert~P. Philipse, and Andrei~V. Petukhov.
\newblock Convectively assembled monolayers of colloidal cubes: Evidence of optimal packings.
\newblock {\em Langmuir} \textbf{2019}, 35(14):4946--4955

% \bibitem{TenNapel2021}
% Daniël N.~ten Napel, Janne-Mieke Meijer, and Andrei~V. Petukhov.
% \newblock The analysis of periodic order in monolayers of colloidal superballs.
% \newblock {\em Applied Sciences} \textbf{2021}, 11(11):5117

\bibitem{Wang2018}
Da~Wang, Michiel Hermes, Ramakrishna Kotni, Yaoting Wu, Nikos Tasios, Yang Liu,
  Bart {De Nijs}, Ernest~B. {Van Der Wee}, Christopher~B. Murray, Marjolein
  Dijkstra, and Alfons {Van Blaaderen}.
\newblock {Interplay between spherical confinement and particle shape on the
  self-assembly of rounded cubes}.
\newblock {\em Nature Communications} \textbf{2018}, 9,2228

\bibitem{Agthe2016}
Michael Agthe, Tom{\'{a}}s~S. Plivelic, Ana Labrador, Lennart Bergstr{\"{o}}m,
  and German Salazar-Alvarez.
\newblock {Following in Real Time the Two-Step Assembly of Nanoparticles into
  Mesocrystals in Levitating Drops}.
\newblock {\em Nano Letters} \textbf{2016}, 16(11):6838--6843

\bibitem{Tang2020}
Yingying Tang, Leyre Gomez, Arnon Lesage, Emanuele Marino, Thomas~E. Kodger,
  Janne-Mieke Meijer, Paul Kolpakov, Jie Meng, Kaibo Zheng, Tom Gregorkiewicz,
  and Peter Schall.
\newblock Highly stable perovskite supercrystals via oil-in-oil templating.
\newblock {\em Nano Letters} \textbf{2020}, 20(8):5997--6004

\bibitem{Sen2014}
D.~Sen, J.~Bahadur, S.~Mazumder, G.~Santoro, S.~Yu, and S.~V. Roth.
\newblock {Probing evaporation induced assembly across a drying colloidal
  droplet using in situ small-angle X-ray scattering at the synchrotron
  source}.
\newblock {\em Soft Matter} \textbf{2014}, 10(10):1621--1627


\bibitem{Marino2018}
Emanuele Marino, Thomas~E. Kodger, Gerard~H. Wegdam, and Peter Schall.
\newblock {Revealing Driving Forces in Quantum Dot Supercrystal Assembly}.
\newblock {\em Advanced Materials} \textbf{2018}, 30(43):1803433

\bibitem{Montanarella2018}
Federico Montanarella, Jaco~J. Geuchies, Tonnishtha Dasgupta, P.~Tim Prins, Carlo van~Overbeek, Rajeev Dattani, Patrick Baesjou, Marjolein Dijkstra, Andrei~V. Petukhov, Alfons van Blaaderen, and Daniel Vanmaekelbergh
\newblock {Crystallization of Nanocrystals in Spherical Confinement Probed by in Situ X-ray Scattering}
\newblock{\em{Nano Letters}} \textbf{{2018}}, {18}({6})

\bibitem{Elkies1991}
N~D Elkies, A~M Odlyzko, and J~A Rush.
\newblock On the packing densities of superballs and other bodies.
\newblock {\em Invent. math} \textbf{1991}, 105:613--639

\bibitem{Chen1998}
S.L.~Chen.
\newblock{Preparation of monosize silica spheres and their crystalline stack}.
\newblock{\em{Colloids and Surfaces A: Physicochemical and Engineering Aspects}} \textbf{{1998}},{142}:{59--63}

\bibitem{Dekker2020b}
F.~Dekker, B.~W.M. Kuipers, González García, R.~Tuinier, and A.~P. Philipse.
\newblock Scattering from colloidal cubic silica shells: Part II, static
  structure factors and osmotic equation of state.
\newblock {\em Journal of Colloid and Interface Science} \textbf{2020}, 571:267--274

\bibitem{Zheng2011}
 Yan Zeng, Stefan Grandner, Cristiano~L.P. Oliveira, Andreas~F. Thünemann, Oskar Paris, Jan~S. Pedersen, Sabine~H.L. Klappb,  and  Regine~von Klitzing.
\newblock {Effect of particle size and Debye length on order parameters of colloidal silica suspensions under confinement}
\newblock {\em{Soft Matter}} \textbf{2011} 7:10899-10909


\bibitem{Pedersen1997}
Jan~Skov Pedersen.
\newblock {Analysis of small-angle scattering data from colloids and polymer
  solutions: Modeling and least-squares fitting}.
\newblock {\em Advances in Colloid and Interface Science} \textbf{1997}, 70(1-3):171--210

\bibitem{Rieker1999}
Thomas Rieker, Aree Hanprasopwattana, Abhaya Datye, and Paul Hubbard.
\newblock Particle size distribution inferred from small-angle x-ray scattering
  and transmission electron microscopy.
\newblock {\em{Langmuir}} \textbf{1999}, 15:638--641

\bibitem{Meijer2013}
Janne~Mieke Meijer, Dmytro~V. Byelov, Laura Rossi, Anatoly Snigirev, Irina
  Snigireva, Albert~P. Philipse, and Andrei~V. Petukhov.
\newblock {Self-assembly of colloidal hematite cubes: A microradian X-ray
  diffraction exploration of sedimentary crystals}.
\newblock {\em Soft Matter} \textbf{2013}, 9(45):10729--10738

\bibitem{Olivero1977}
J.~J. Olivero and R.~L. Longbothum.
\newblock {Empirical fits to the Voigt line width: A brief review}.
\newblock {\em Journal of Quantitative Spectroscopy and Radiative Transfer} \textbf{1977},
  17(2):233--236

\bibitem{Patterson1939}
A.~L. Patterson.
\newblock {The scherrer formula for X-ray particle size determination}.
\newblock {\em Physical Review} \textbf{1939}, 56(10):978--982

\bibitem{Petukhov2003}
A.~V. Petukhov, I.~P. Dolbnya, D.~G.A.L. Aarts, G.~J. Vroege, and H.~N.W.
  Lekkerkerker.
\newblock Bragg rods and multiple x-ray scattering in random-stacking colloidal
  crystals.
\newblock {\em Physical Review Letters} \textbf{2003}, 90(2):028304 

\bibitem{Kralchevsky1994}
Peter~A. Kralchevsky, and Kuniaki Nagayama
\newblock Capillary forces between colloidal particles
\newblock {\em Langmuir}\textbf{1994}, 10(1):23-36

\bibitem{Sugimoto1992}
Tadao Sugimoto and Kazuo Sakata.
\newblock {Preparation of monodisperse pseudocubic $\alpha$-Fe2O3 particles
  from condensed ferric hydroxide gel}.
\newblock {\em Journal of Colloid and Interface Science} \textbf{1992}, 152(2):587--590

\bibitem{Stober1968}
Werner St{\"{o}}ber, Arthur Fink, and Ernst Bohn.
\newblock {Controlled growth of monodisperse silica spheres in the micron size
  range}.
\newblock {\em Journal of Colloid and Interface Science} \textbf{1968}, 26(1):62--69

\bibitem{Deng2012}
Xu~Deng, Lena Mammen, Hans~J{\"{u}}rgen Butt, and Doris Vollmer.
\newblock {Candle soot as a template for a transparent robust superamphiphobic
  coating}.
\newblock {\em Science} \textbf{2012}, 335(6064):67-70

\bibitem{Bras2003}
W.~Bras, I.P. Dolbnya, D.~Detollenaere, R.~van Tol, M.~Malfois, G.N. Greaves,
  A.J. Ryan, and E.~Heeley.
\newblock {Recent experiments on a small-angle/wide-angle X-ray scattering beam
  line at the ESRF}.
\newblock {\em Journal of Applied Crystallography} \textbf{2003}, 36(3 Part 1):791-794

\bibitem{Snigirev1996}
A~Snigirev, V~Kohn, I~Snigireva, and B~Lengeler.
\newblock {A compound refractive lens for focusing high-energy X-rays}.
\newblock {\em {Nature}} \textbf{{1996}}, {384}({6604}):{49-51} 

\bibitem{Wang2021b}
J. Wang, J. Schwenger, A. Strobel, P. Feldner, P. Herre, S. Romeis, W. Peukert, B. Merle, and N. Vogel.
\newblock {Mechanics of colloidal supraparticles under compression}.
\newblock{\em{Science Advances}}\textbf{2021}, 7(42):eabj0954

\bibitem{Graf2003}
Christina Graf, Dirk L. J. Vossen, Arnout Imhof, and Alfons van Blaaderen.
\newblock {A General Method To Coat Colloidal Particles with Silica}.
\newblock {\em{Langmuir}} \textbf{{2003}},{19} ({17}):{6693–6700}

\bibitem{Petukhov2015}
Andrei~V. Petukhov, Janne-Mieke Meijer, Gert Jan Vroege
\newblock{Particle shape effects in colloidal crystals and colloidal liquid crystals: Small-angle X-ray scattering studies with microradian resolution}.
\newblock{\em{Current Opinion in Colloid Interface Science}} \textbf{2015}, 20(4):272--281



\end{thebibliography}
\end{document}

\title{Supplementary Information: Self-assembly of Colloidal Superballs Under Spherical Confinement of a Drying Droplet}

\author{Sarah N.\ \surname{Schyck}*}
\affiliation{Delft University of Technology,Department of Chemical Engineering, 2629 HZ Delft, The Netherlands}

\author{Janne-Mieke\ \surname{Meijer}*}
\affiliation{Eindhoven University of Technology, Department of Applied Physics and Insitute for Complex Molecular Systems, Eindhoven, The Netherlands}

\author{Lucia\ \surname{Baldauf}}
\affiliation{University of Amsterdam, Institute of Physics, Science Park 904, Amsterdam, The Netherlands}
\affiliation{Present address: Delft University of Technology,Department of Bionanoscience, Kavli Institute of Nanoscience, Delft 2629 HZ Delft, The Netherlands}

\author{Peter\ \surname{Schall}}
\affiliation{University of Amsterdam, Institute of Physics, Science Park 904, Amsterdam, The Netherlands}

\author{Andrei V.\ \surname{Petukhov}}
\affiliation{Utrecht University, Debye Institute for Nanomaterials Science, Utrecht, the Netherlands}
\affiliation{Eindhoven University of Technology, Department of Chemical Engineering and Chemistry, Eindhoven, the Netherlands}

\author{Laura\ \surname{Rossi}}
\email{L.Rossi@tudelft.nl}
\affiliation{Delft University of Technology,Department of Chemical Engineering, 2629 HZ Delft, The Netherlands}

\date{\today}

\maketitle

%%%%%%%%%%%%

%
%%%%%%%%%%%
%
\begin{table*}[h!]
\centering
\caption{A table of all samples and their parameters used in this work. Solid silica spheres are denoted by the prefix {\em{S}} while hollow silica superballs are denoted by the {\em{HSB}} prefix in the sample name. The outer {\em{m}} value denotes the shell shape while the inner {\em{m}} denotes the core shape. The thickness, {\em{t}}, is in nm along with half the distance between two flat faces, {\em{a}}, or the radius in the case of spherical particles. }
\begin{tabular}[t]{l>{\centering}m{0.2\linewidth}m{0.15\linewidth}m{0.15\linewidth}c}
\toprule
Name & Outer \em{m} & Inner \em{m} & \em{t}(nm) & \em{a}(nm)\\
\midrule
S1      &  2 & -- & -- & 233\\
S2& 2 & -- & -- & 206\\
\midrule
HSB1 & 2.6 & 3.65 & 220 & 512\\
HSB2 &2.8&3.65&187&446\\
HSB3&3.1&3.65&56&315\\
HSB4&3.3&3.9&111&413\\
\bottomrule
\end{tabular}
\end{table*}%

%%%%%%%%%%%%

\begin{figure*}[b]
    \centering
    \includegraphics[width=0.95\textwidth]{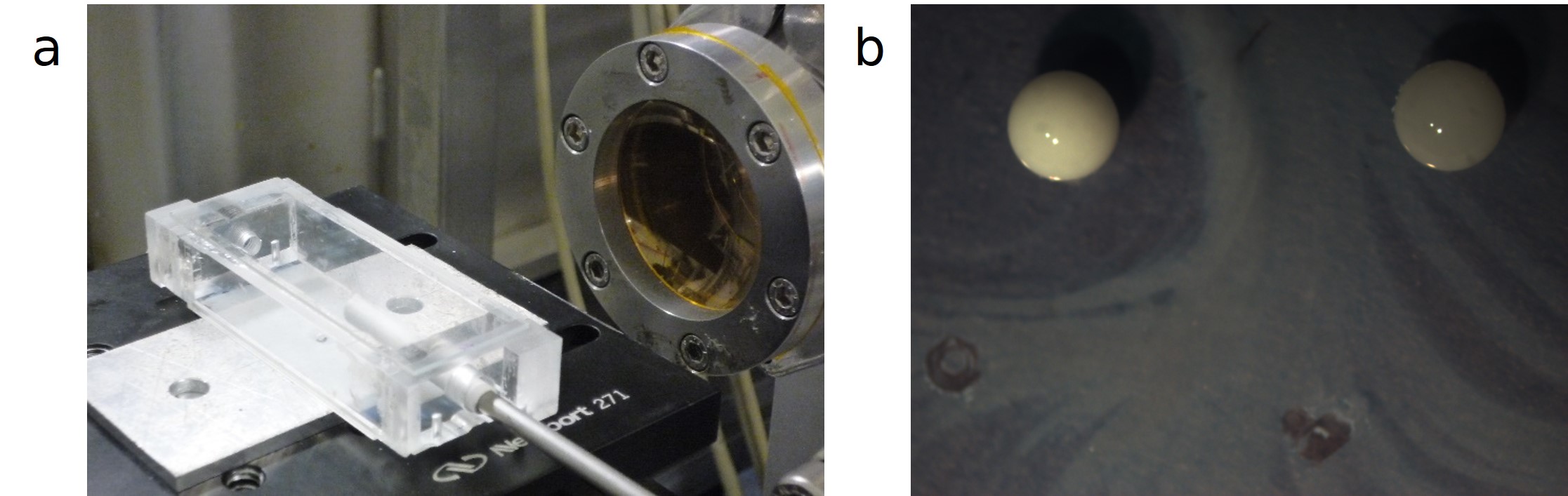}
    \caption{Images of the experimental set-up. (a) A photograph of the custom humidity chamber with hydrophobic plate aligned with the X-ray path for SAXS measurements, and (b) An image of two dispersion droplets of silica spheres (left) and hollow superballs (right) on a superhydrophobic plate. The droplets have a radius of approximately 1-2 mm.}
    \label{fig:Exp}
\end{figure*}

\clearpage
%%%%%%%%%%%%

\begin{figure*}[t]
    \centering
    \includegraphics[width=0.95\textwidth]{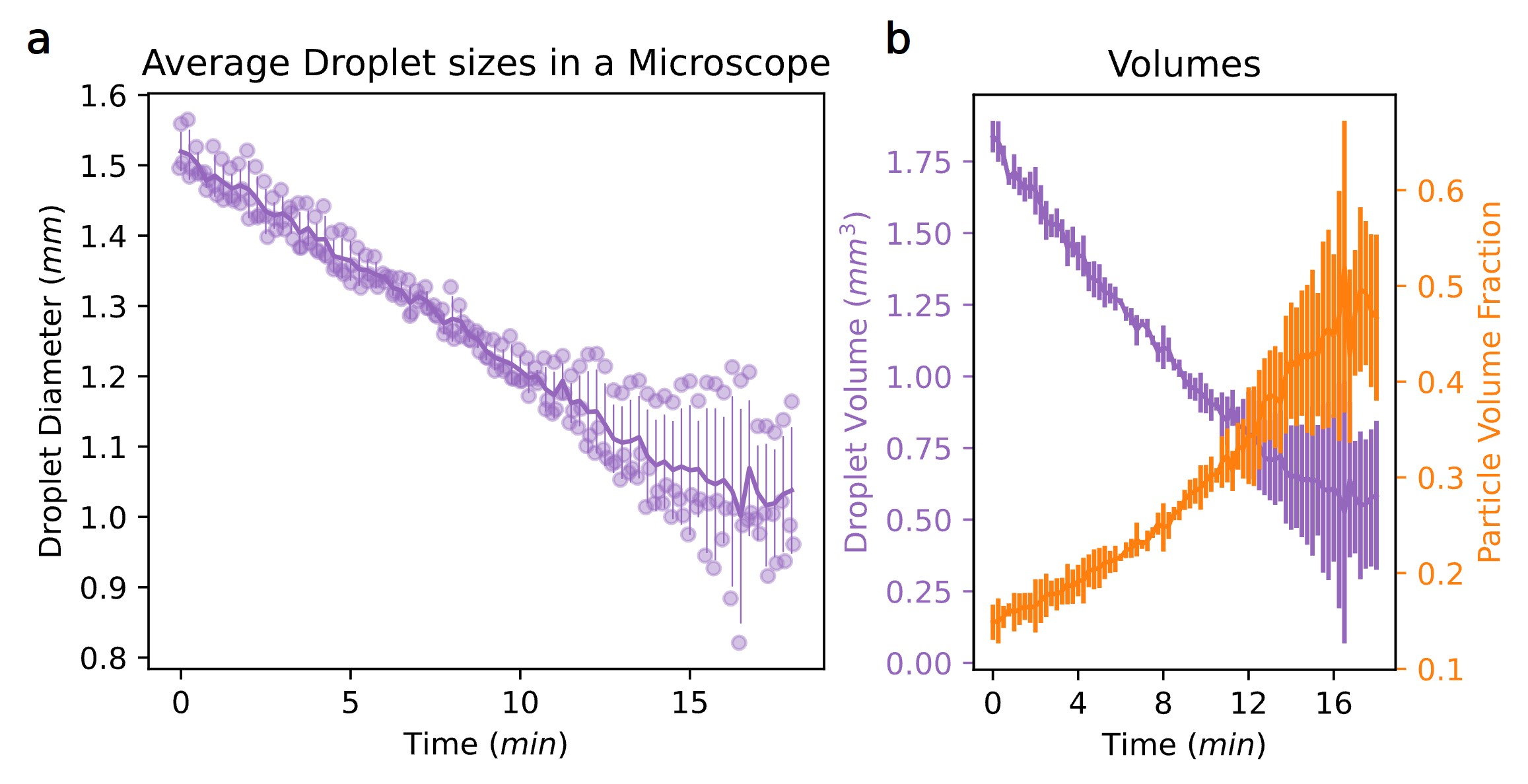}
    \caption{(a) Average droplet size over time for six droplets drying inside the humidity chamber under a microscope camera (purple line) with the standard deviation as the error overlayed with the individual time points of the varying droplets (purple circles). As the droplets dry, the standard deviation in sizes becomes larger because not all droplets reached the same final diameter. (b) The average droplet volume (purple line) calculated from the average droplet diameter in (a) is plotted with the expected volume fraction of superball particles (orange line) when starting from a 15v\% droplet. Analyzed images were collected with a Zeiss Axio Imager.A1 upright microscope at low magnifications. We note that microscope experiments where completed with separate droplets apart from the x-ray experiments. }
    \label{fig:Exp}
\end{figure*}

\clearpage
%%%%%%%%%%%%

%%%%%%%%%%%%

\begin{figure*}[t]
    \centering
    \includegraphics[width=0.95\textwidth]{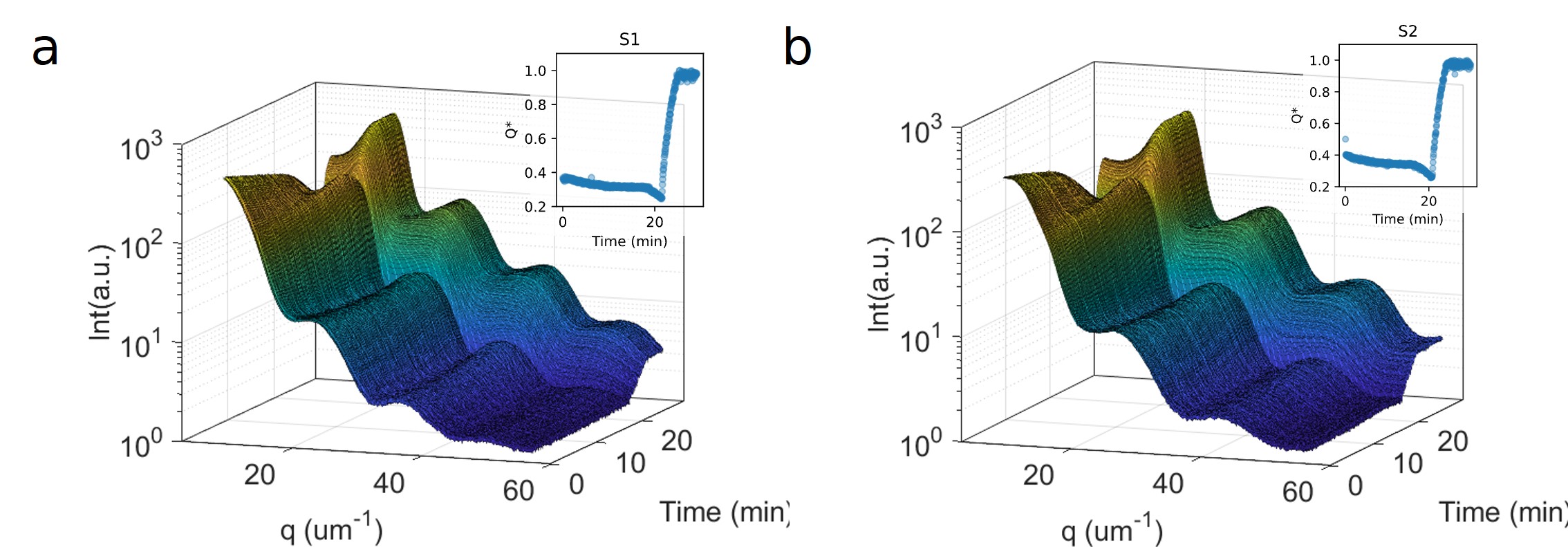}
    \caption{Time resolved scattering profiles of both sphere experiments (a) S1 and (b) S2 for the whole-time range. The insets depict the normalized scattering power, Q*, that corresponds with the experimental scattering profiles. }
    \label{fig:Exp}
\end{figure*}

%%%%%%%%%%%%

\begin{figure*}
    \centering
    \includegraphics[width=0.95\textwidth]{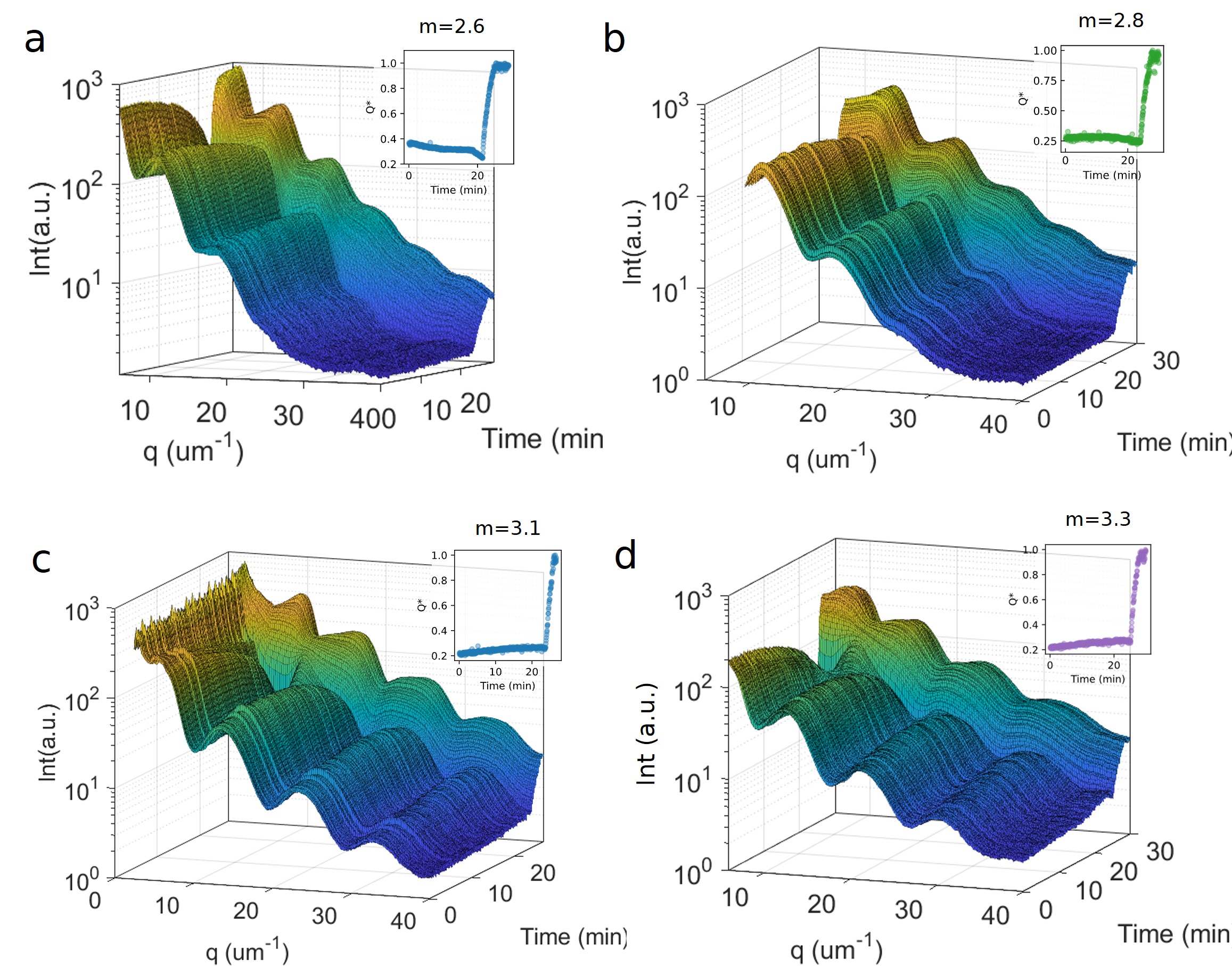}
    \caption{Time resolved scattering profiles of hollow superball dispersion droplets with increasing m value; (a) HSB1, (b) HSB2, (c) HSB3, and (d) HSB4. The insets depict the normalized scattering power, Q*, that corresponds with the experimental scattering profiles. }
    \label{fig:Exp}
\end{figure*}

\clearpage

%%%%%%%%%%%%
%%%%%%%%%%%%

\begin{figure*}[t]
    \centering
    \includegraphics[width=0.95\textwidth]{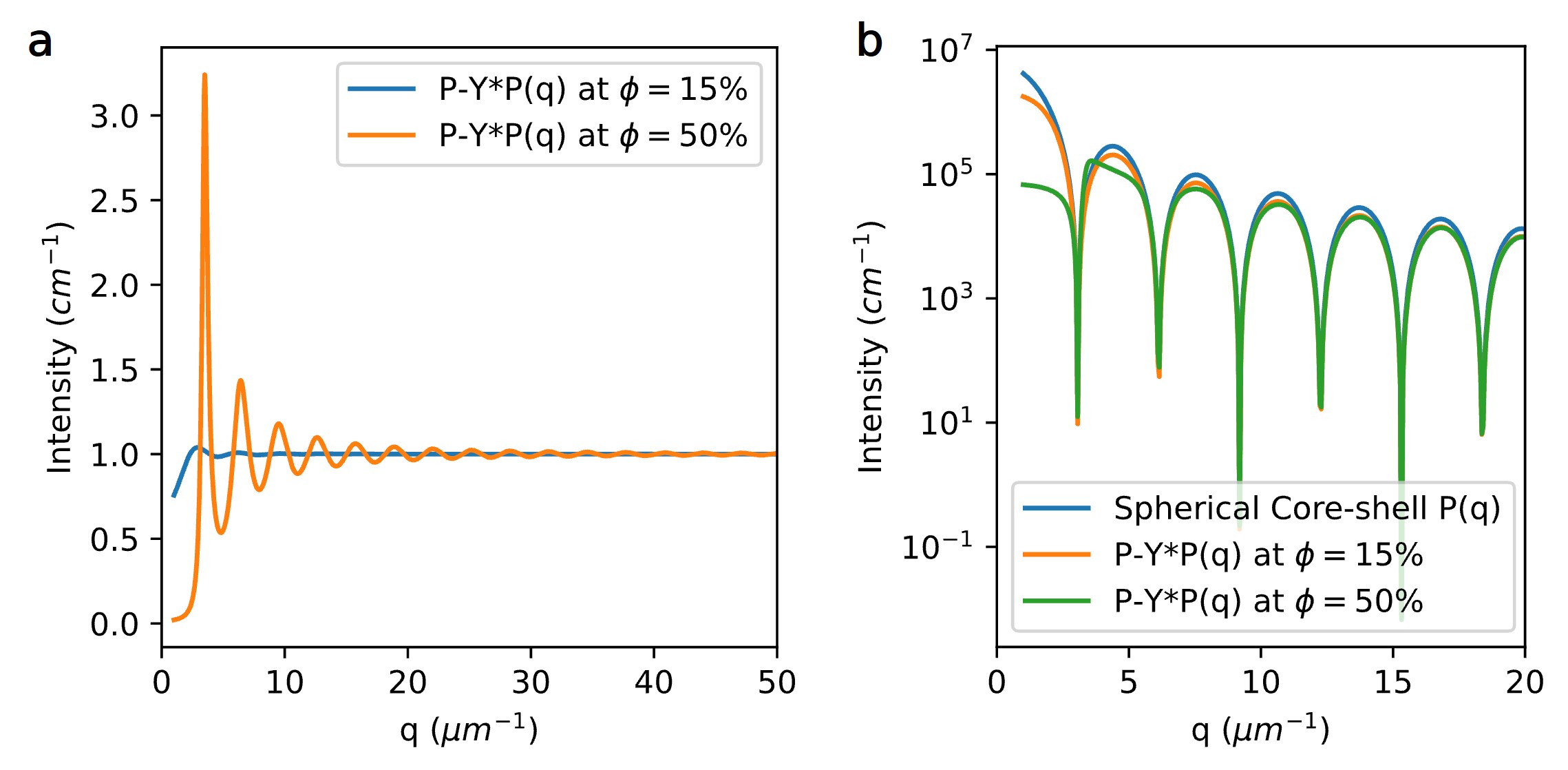}
    \caption{(a) The calculated Percus-Yevick structure factor at a particle volume fraction of $\phi = 15$\% (blue line) and at $\phi = 50$\% (orange line). The Percus-Yevick S(q) is generally used for hard particle interactions, and we have calculated for our starting $\phi$. (b) Here, we use the model form factor of a core-shell spherical particle with a diameter of $1$ $\mu$m defined as $P(q)= [\frac{3}{V_{s}}(V_{c}(\rho_{c}-\rho_{s})\frac{\sin{qr_c}-qr_c\cos{qr_c}}{(qr_c)^3}+V_s(\rho_s-\rho_{solv})\frac{\sin{qr_s}-qr_s\cos{qr_s}}{(qr_s)^3})] $ where $V_s$ is the volume of the shell and core, $V_c$ is the volume of the core, $r_s$ is the radius of the particle, $r_c$ is the radius of the core, $\rho_s$ is the scattering length density (SLD) of the core, $\rho_s$ is the SLD of the shell, and $\rho_{solv}$ is the SLD of the solvent (blue line). We multiply the calculated Percus-Yevick structure factor by the $P(q)$  at $\phi=15$\% (orange line) and $\phi=50$\% (green line). From the curves, we can conclude that while at $15$\% two small $S(q)$ peaks appear around $q=3.5$ and $q=6$ $\mu m^{-1}$, the influence on the overall $I(q)$ is negligible for the experimentally accessible $q$-regime of $q=7-40$ $\mu m^{-1}$. }
    \label{fig:Exp}
\end{figure*}

\clearpage
%%%%%%%%%%%%

\begin{figure*}
    \centering
    \includegraphics[width=0.90\textwidth]{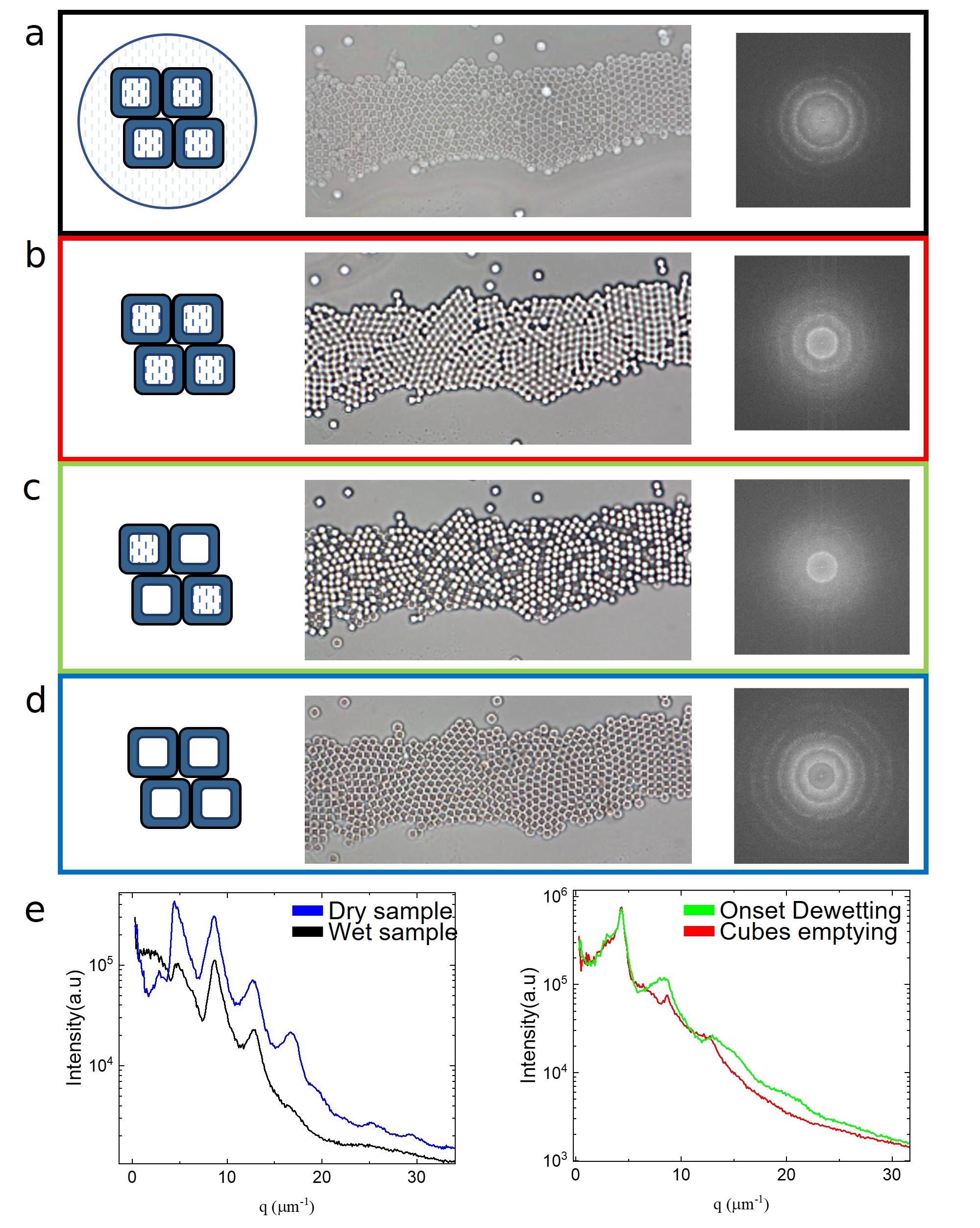}
    \caption{Overview of the drying process directly visualized by optical microscopy of a hollow superball monolayer. For each stage of dewetting, a schematic, optical microscopy images, and corresponding FFT image is provided for the (A) fully wetted monolayer, (B) onset of dewetting, (C) superballs emptying, and (D) fully dry monolayer. (E) Compares the 1D radially averages of the FFT images for each stage of dewetting. The optical microscopy experiment was done on a Nikon Eclipse inverted microscope, and the sample was recorded over 300s with a frame rate of 2 frames per second. FFT of the monolayers was done via ImageJ and 1D patterns were plotted for comparison.}
    \label{fig:Exp}
\end{figure*}
%%%%%%%%
 
%%%%%%%%%%%%

\begin{figure*}[t]
    \centering
    \includegraphics[width=0.9\textwidth]{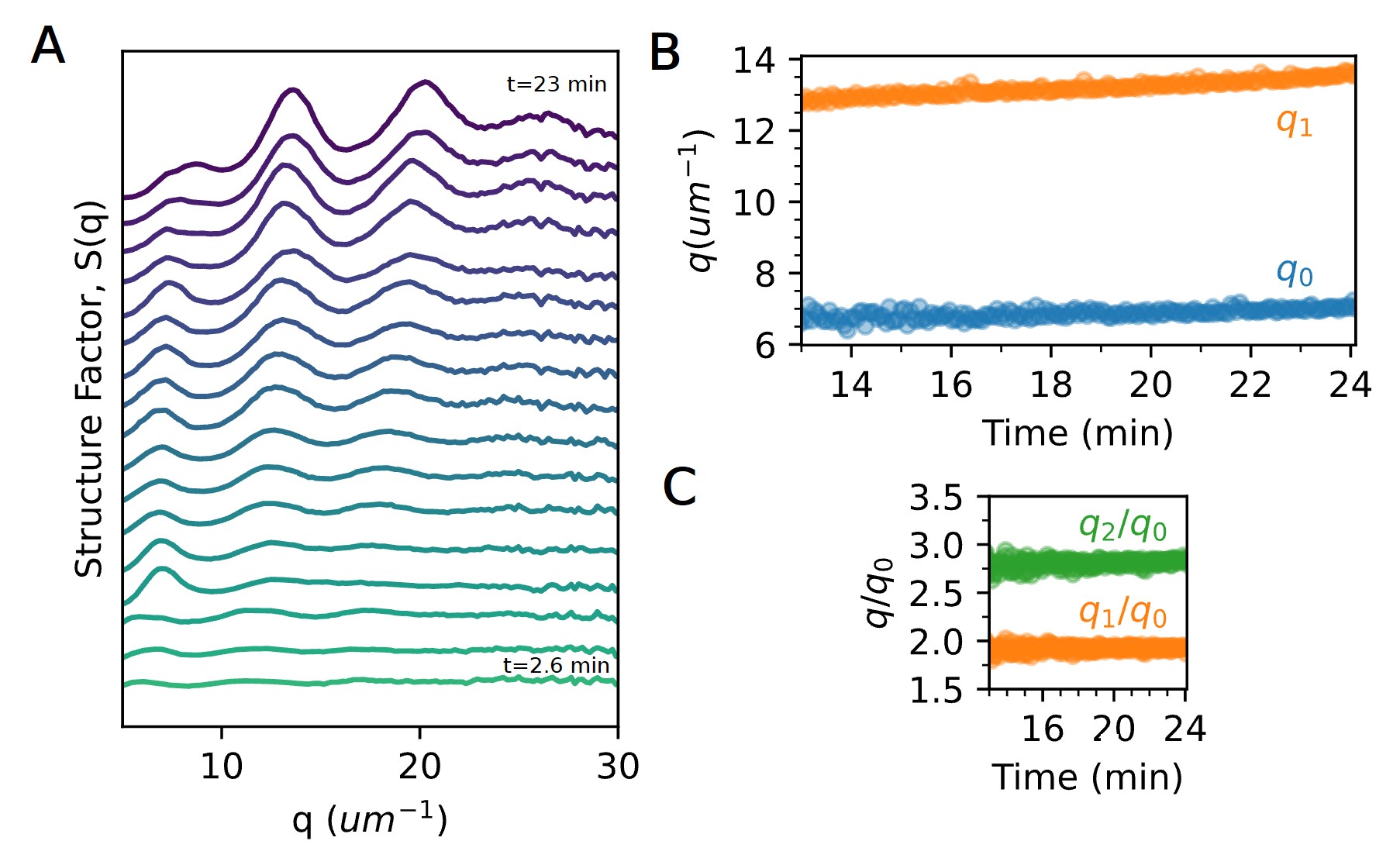}
    \caption{(a) Time resolved scattering profile for $m=2.6$ where the time step between curves is 1.2 min. (b) Peak position of $q_0$ and $q_1$ over time - excluding the shoulder peak that emerges from the $q_0$. (c) The ratio of $q_1/q_0$ and $q_2/q_0$ as the droplet dries. The ratio remains relatively constant as the droplet dries.}
    \label{fig:Exp}
\end{figure*}

\clearpage
%%%%%%%%%%%%

%%%%%%%%%%%%

\begin{figure*}
    \centering
    \includegraphics[width=0.95\textwidth]{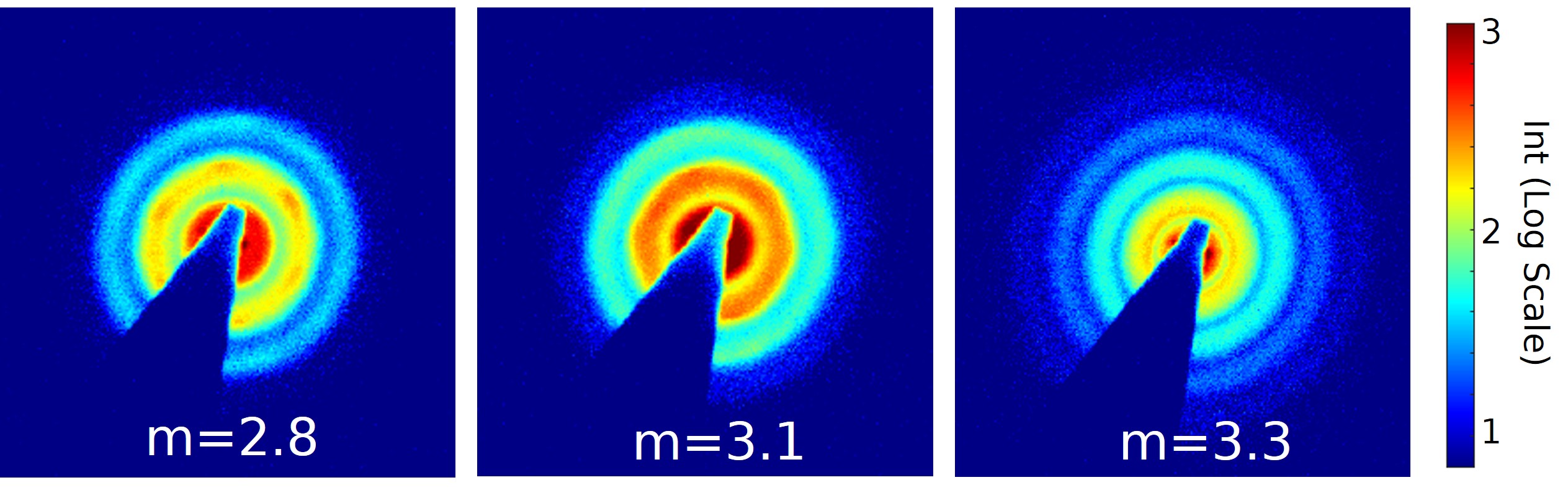}
    \caption{2D SAXS images of samples m=2.8, m=3.1, and m=3.3 before the dewetting transition. }
    \label{fig:Exp}
\end{figure*}

%%%%%%%%%%%%

\begin{figure*}
    \centering
    \includegraphics[width=0.95\textwidth]{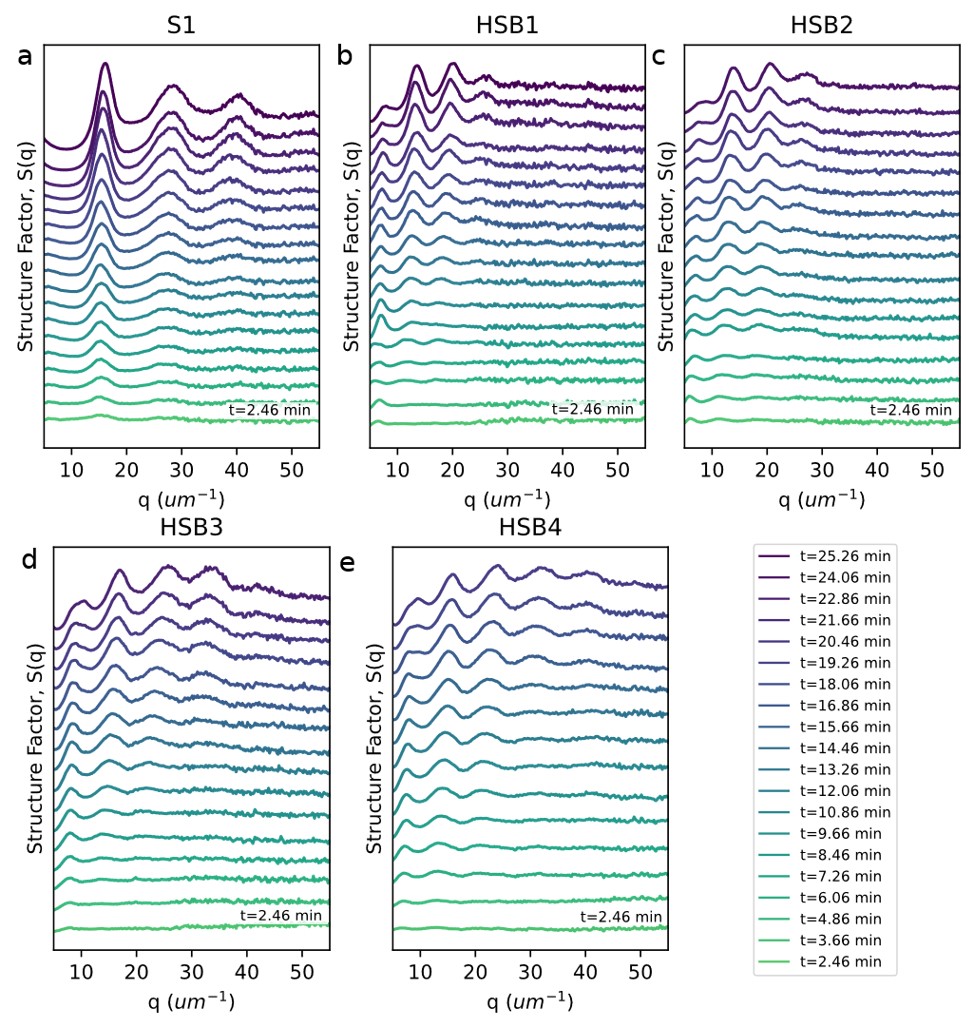}
    \caption{Selected structure factor patterns are stacked in these plots until the de-wetting point which was extracted from the normalized scattering power, Q*. The experiment (a) for spheres, S1, is shown until 22.86 minutes. For hollow superballs, (b) HSB1, (c) HSB2, (d) HSB3, and (e) HSB4 is plotted up until 24.06, 25.26, 24.06, and 25.26 minutes, respectively. The time between the stacked patterns in all panels has a consistent gap of 1.2 minutes between patterns (as shown in the legend in the last panel).}
    \label{fig:Exp}
\end{figure*}

\clearpage

%%%%%%%%%%%%
\begin{figure*}[ht!]
  \includegraphics[width=0.95\textwidth]{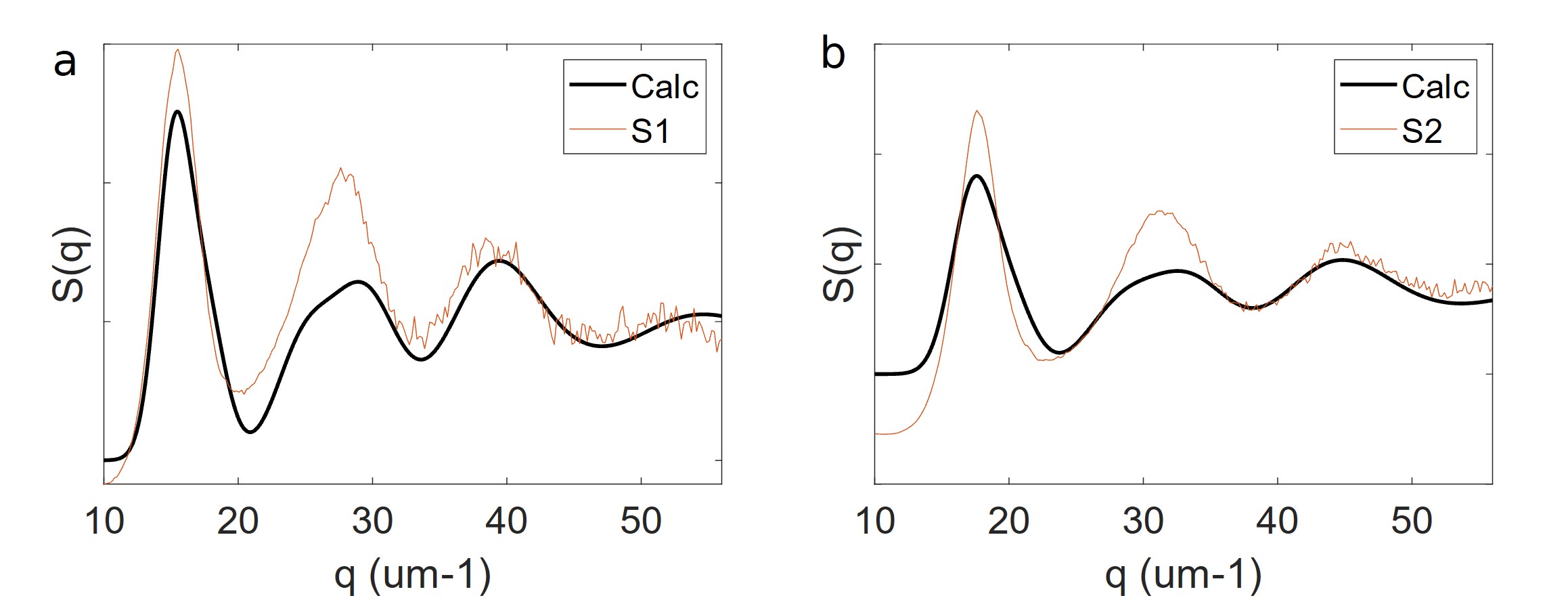}
    \caption{The structure factor of (a) S1 and (b) S2 before dewetting at 22.86 minutes (orange line). The proposed para-cystalline model (thick black line) has the same distortion factor, $\sigma = 0.11$, for (a) and (b).}
    \label{fig:Exp}      
\end{figure*}
\clearpage
%%%%%%%%%%%%
\begin{figure*}
    \centering
    \includegraphics[width=0.5\textwidth]{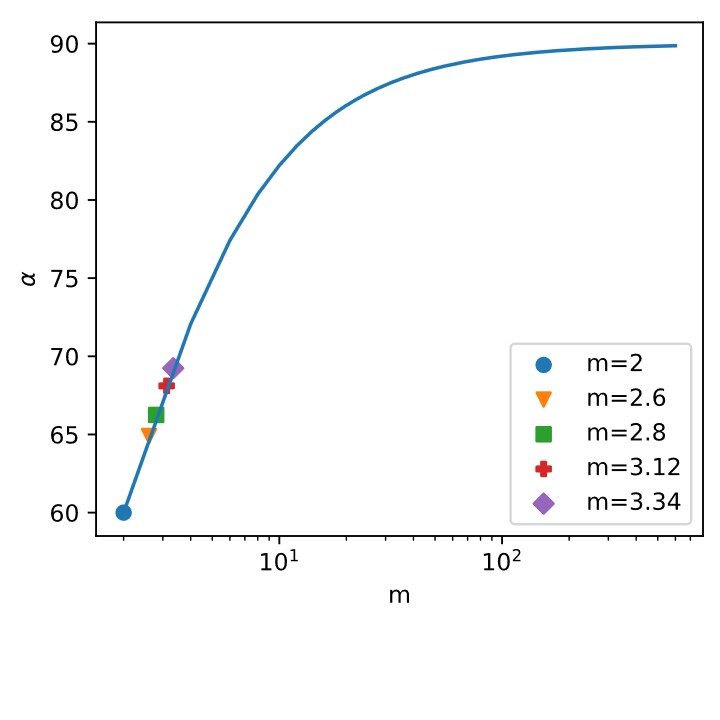}
    \caption{Model line of $\alpha$ vs the $m$ value. Markers are for the experimental $m$ values. For calculations involved, see the main text.}
    \label{fig:Exp}
\end{figure*}

\clearpage
%%%%%%%%%%%%%

\section{Discussion over the Porod Invariant}
We denote $\phi_0$ as the volume fraction occupied by particles and $\phi_1$ as the ratio of the void volume relative particle volume. The volume outside the particles $(1-\phi_0)$ is occupied by water in the wet state and by air in the partial and complete drying cases. We further assume that the voids, $\phi_0\cdot \phi_1$,  inside the particles are filled with water in the wet and partially dry state and with air in the completely dry state. The remaining volume $\phi_0\cdot (1-\phi_1)$ is silica and we assume its density does not change much in going from the wet to dry states.

The realistic value of $\phi_0$ is  $\sim 0.6$ at the moment when air enters the droplet, and $\phi_1 = 0.70-0.85$ based on EM images. The scattering length density contrast between silica and air is a factor of $1.8-2$ times higher than that between silica and water. With these values one expects that the Porod invariant, $Q$, should increase $3.4-4$ times by going from the wet to completely dry state while the partially wet state should have the $Q$ value close to that in the completely dry state within the range of the numerical values mentioned above.

\section{Description of the Dewetting Process}
Figure S4 depicts the drying process in a 2D system made of a superball monolayer. As the system dries, we can clearly identify the distinct phases we have seen in the SAXS patterns. Initially, the monolayer is fully wetted where the particle cores and surrounding medium are water. This creates a defined contrast between the particles’ shell and makes them easily visible. As the water in the surrounding medium evaporates, the particles’ core remains filled. In B, we can no longer visibly see the shell, but rather the entire particle. When the water empties out, we see how each individual particle changes rather than all at once. Eventually, all of the water empties from the superball cores and we have a better contrast in the images due to the complete change in medium.

 We collected fast Fourier transforms (FFT) of the monolayers for each phase of dewetting. By directly observing and radially averaging them, we can see how each phase changes. For the fully wet and fully dry monolayer, the patterns are remarkable similar to each other and are mostly dominated by the $P(q)$ peaks. As the cubes dewet, we see the patterns change where $P(q)$ completely disappears as the particles empty, and there is only a small contribution from the $S(q)$ peaks. By visualizing a simpler system, we better identify how the changing contrast and environment relates to changes in the SAXS patterns even though the physical structure doesn’t change further.

\section{Para-crystalline Model}
For the model of a para-crystalline structure, we modify the description from Förster {\em{et al.}}\cite{Forster2007} for a similar type of structure. We describe $I(q)$ as follows
\begin{equation*}
	I(q) \propto P(q) \cdot [1+(Z(q)-1)\cdot G(q)]
	\end{equation*}
where
\begin{equation*}
	G(q) = \exp \Big( -\frac{q^2 \langle u \rangle^2}{3} \Big)
	\end{equation*}
is the usual Debye-Waller factor with the mean square displacement of the lattice points, $\langle u \rangle^2$, and
\begin{equation*}
	Z(q)=\sum_{hkl}\frac{2M}{\pi\sigma q^2 q_{hkl}}\exp \Big[-\frac{4(q-q_{hkl})^2}{\pi(\sigma q_{hkl})^2}\Big]
\end{equation*}
Here, $q_{hkl}$, is the reciprocal lattice vector for the $hkl$ reflection, $\sigma$ is the distortion factor, and $M$ is the multiplicity.  For simplicity, we assume there is no Debye-Waller factor, i.e., $G\left(q\right)=1$. In this version, we also assume that the peak width increases for higher $q_{hkl}$ values. For FCC, the reciprocal lattice vector is defined as $q_{hkl}=\frac{2\pi}{a}\sqrt{h^2+k^2+l^2}$. Fig S5 shows this model for the two experiments involving spheres (S1 and S2). Since the distortion factor is kept constant ($\sigma=\ 0.11$), the model differences arises from differences in the size of the spherical particles where the radii for S1 (r=250nm) and S2 (r=207nm) are different.

\pagebreak